\documentclass{aa}
 
\usepackage{graphicx}
\usepackage{txfonts}            
\usepackage{amsmath}
\usepackage{subcaption}         
\usepackage[dvipsnames]{xcolor}
\usepackage[normalem]{ulem}

\usepackage{hyperref}
\hypersetup{
    colorlinks=true,
    linkcolor=blue,    
    citecolor=blue,    
    urlcolor=blue,     
    linkbordercolor={1 1 1}
}
 
\begin{document}

\title{Chemical evolution of the Milky Way disc with radial gas flows: \\ a Lagrangian approach}

\author{Fiorenzo Vincenzo}
 
\institute{Dipartimento di Fisica e Astronomia ``Ettore Majorana'', Universit\`a degli Studi di Catania, Via S. Sofia 64, 95123 Catania, Italy\\ 
\email{fiorenzo.vincenzo@dfa.unict.it}}
 \date{Received 29 June 2026 / Accepted 14 September 2026}

\abstract
{Chemical abundance patterns result from the interplay between gas accretion, star formation, and radial mixing of gas and stars. Disentangling these processes is crucial to recover the mechanisms shaping the formation and evolution of galaxies.}
{We model the chemical evolution of the Galactic disc in the presence of radial gas flows, to assess their impact on the [O/Fe]-[Fe/H] abundance patterns and on the radial gradients of [Fe/H] and [O/H].}
{We develop fast, semi-analytic solutions for the gas surface mass density and the abundances of $\alpha$-elements and iron, accounting for radial gas flows and chemical enrichment from core-collapse and Type~Ia supernovae. The model follows a Lagrangian approach, using the method of characteristics, reducing the solutions to one-dimensional integrals. We apply our model to the Milky Way disc assuming a two-infall scenario.}
{When radial gas flows are present, the chemical abundances of the gas at a given radius result from its whole inward journey in the disc, reflecting the star formation and accretion experienced at every radius it crossed. The integrated stellar mass along the characteristic is lower than the local value by up to an order of magnitude at $v \simeq 1.5~\mathrm{km\,s^{-1}}$. Models with mild flows of $v \simeq 1.5~\mathrm{km\,s^{-1}}$ reproduce simultaneously the observed [O/Fe]-[Fe/H] distribution across the disc, the present-day stellar surface-density profile, and the [Fe/H] and [O/H] gradients, improving also the agreement with the observed age-abundance relations. The stellar mass formed per Type~Ia supernova sets the [O/Fe] ratio and departs from its in-situ value by up to $\sim50$ per cent, making the $\alpha$-enhancement the quantity on which radial flows leave their strongest signature.}
{Following the gas along its trajectory is essential to recover the correct enrichment history even for models with mild radial gas inflows. The code is made publicly available.}
 

\titlerunning{Radial gas flows and chemical evolution}
\authorrunning{Fiorenzo Vincenzo}

\maketitle
 \nolinenumbers
 
\section{Introduction}

Large spectroscopic surveys like SDSS-APOGEE, which targeted and provided spectra for $\approx 7 \times 10^{5}$ stars in our Galaxy \citep{Majewski2017, Abdurrouf2022}, Gaia-ESO with its $\approx 10^{5}$ stars \citep{Hayden2018, Randich2022}, GALAH with $\approx 9\times 10^{5}$ stars \citep{Buder2018, Hayden2020, Buder2025}, and LAMOST with $\approx 1.1 \times 10^{7}$ stars \citep{Luo2015, Xiang2019} have completely reshaped our view of the formation and evolution of our Galaxy over the last decade, by revealing how the measured chemical abundances correlate with one another for large samples of stars across a wide range of distances in the Milky Way (MW) \citep{Freeman2002, Weinberg2019, Weinberg2022, Deason2024}. Fortuitously, these surveys came of age alongside Gaia, whose precise parallaxes and proper motions for $\approx 1.46 \times 10^{9}$ stars \citep{Gaia2023} could be coupled with the chemical-abundance measurements: without Gaia, this revolution would not have had the same reach. In the near future, upcoming surveys like MOONS with $\approx 10^{6}$ stars \citep{Cirasuolo2014, Gonzalez2020}, 4MOST with $\approx 10^{7}$ stars \citep{Chiappini2019, Bensby2019, deJong2019}, and WEAVE with $\approx 10^{6}$ stars \citep{Jin2024} promise a further revolution in Galactic Archaeology studies and related subdisciplines.

The way in which stellar chemical abundances are distributed across the Galactic disc results from the synergy of several processes acting together; ultimately, they directly sample how short- and long-lived stars of previous generations released their nucleosynthetic products relative to the past star formation and gas accretion history \citep[e.g.,][]{Tinsley1980, Matteucci1986}. Examples of processes explored in chemical evolution studies for establishing radial chemical abundance gradients in the Galaxy disc are the inside-out growth of the disc in combination with a radial dependence of the star formation efficiency \citep{Matteucci1989, Molla2016, Johnson2021}; in addition to them, other works have also discussed the importance of stellar migration \citep{Schoenrich2009, Spitoni2015, Kubryk2015, Johnson2021, Sharma2021}, concluding that the effect of radial mixing of stars tends to flatten the elemental abundance gradients as a function of time. 

Another mechanism for the development of radial abundance gradients in the Galaxy disc that has been widely discussed in the literature is given by radial gas flows. According to the main proposed scenario, radial gas flows stem from an angular momentum difference between the infalling gas on the disc, that carries lower average angular momentum, and the rotating gas on the disc, with higher average angular momentum; the mixing between the two drives a net inward motion with speeds of at most a few $\mathrm{km\,s^{-1}}$ \citep{Lacey1985, Bilitewski2012}. In particular, \citet{Bilitewski2012} find that the accreting material must rotate at $70$--$75$ per cent of the disc circular velocity to reproduce the observed [Fe/H] gradient in the Galaxy disc.

It has been shown by several chemical evolution models in the past that the net effect of radial gas inflows is to lower the gas metallicity and to steepen the radial abundance gradients \citep{Portinari2000, Spitoni2011, Bilitewski2012, Spitoni2013, Palla2020}. One of the most detailed studies is that of \citet{Palla2020}, who could reproduce the MW abundance gradients with a state-of-the-art chemical evolution model by assuming a two-infall scenario for the mass assembly of the MW thick and thin disc alongside radial gas flows with constant velocity, a radially varying star formation efficiency, and inside-out mass growth. Another relevant theoretical study in the literature is that of \citet{Pezzulli2016}, who showed that even a small angular-momentum difference inevitably drives radial gas flows, developing an original mathematical formalism with general analytic solutions for the gas accretion profiles, radial mass fluxes, and abundance gradients as a function of the angular momentum of the accreting material; crucially, they solved the partial differential equation for the metallicity evolution in the presence of radial gas flows by making use of the method of characteristics. Recently, \citet{Johnson2025} reached a consistent conclusion by developing multi-zone chemical evolution models, finding that radial flows systematically lower the metallicity and steepen the radial abundance gradients.

A distinctive feature of the chemical abundance pattern of the MW disc is the so-called bimodality between the high-[$\alpha$/Fe] thick-disc stars and the low-[$\alpha$/Fe] thin-disc stars in the [$\alpha$/Fe]--[Fe/H] plane \citep{Fuhrmann1998, Bensby2003, Hayden2015}. A scenario that has been postulated to explain this observed bimodality in [$\alpha$/Fe] is the so-called two-infall scenario of \citet[see also \citealt{Spitoni2019, Spitoni2020, Spitoni2021, Palla2020}]{Chiappini1997}, in which the disc of our Galaxy assembled through two distinct accretion episodes: \textit{(i)} a short, early one that formed the thick disc, and \textit{(ii)} a longer, later one that formed the thin disc, with a period of reduced star formation in between. This framework has been shown to reproduce the observed [$\alpha$/Fe]--[Fe/H] distribution across the disc by SDSS-APOGEE alongside the local surface densities of stars and gas \citep{Spitoni2019, Spitoni2020, Palla2020, Spitoni2021}. The other main competing mechanism that has been advanced to explain the observed [$\alpha$/Fe] bimodality is given by radial stellar migrations (see \citealt{Schoenrich2009, Sharma2021}, but also \citealt{Johnson2021} for a different perspective in which stellar migration is coupled consistently with chemical evolution). Nevertheless, the advent of large stellar age catalogues has recently introduced new tensions; in particular, \citet{dubay2026} showed that, while the two-infall scenario can explain the observed [$\alpha$/Fe]-bimodality, it struggles to reproduce the age--abundance structure of the MW disc, as the data show a metallicity that stays nearly constant as a function of stellar age for much of the disc lifetime, whereas two-infall models generically predict a substantial dilution event at the onset of the thin disc, leaving a strong signature on the predicted chemical abundances.

In this work we build on the characteristic-based approach of \citet{Pezzulli2016}, extending it in two directions that, together, allow us to confront the model directly with the resolved [$\alpha$/Fe]--[Fe/H] abundance pattern of the MW disc. In our work we use O as a proxy for the $\alpha$-elements. First, our chemical evolution model solves both the metallicity and the gas surface density along the exact same characteristics. Rather than prescribing a gas profile and decomposing the implied mass flux, we obtain the gas surface mass density as an explicit integral solution of its evolution equation, so that the structural and chemical evolution of the disc are followed self-consistently within a single mathematical formalism. Second, we follow the evolution of the chemical abundances of both $\alpha$-elements, produced by short-lived massive stars exploding as core-collapse supernovae (SNe), and iron-peak elements, mainly produced by exploding white dwarfs in binary systems as Type Ia SNe on a distribution of delay times from the star formation event. This allows our model to predict an [$\alpha$/Fe]--[Fe/H] chemical abundance pattern, and not just the metallicity gradient. Because iron is largely released by Type~Ia SNe with a delay relative to the prompt production of oxygen by core-collapse SNe, the gas reaching a given radius from the outer disc carries the imprint of fewer Type~Ia SNe, so that radial gas flows raise [O/Fe] in a way that a single-element treatment with instantaneous recycling cannot capture. The $\alpha$-enhancement thus becomes a sensitive diagnostic of the inflow velocity, complementary to and independent of the abundance gradient slope.

A simple idea motivates our Lagrangian approach. When radial gas flows are present, the gas that we observe at a given Galactocentric radius today did not generally form there: it was carried inward from larger radii, and along the way it was enriched by the star formation and gas accretion that took place at each radius it crossed. The chemical abundances we measure at a given radius today are therefore not set by what happened at that radius alone, as a simple one-zone model would assume, but by the whole history of the gas as it flowed inward. Following each gas parcel along its trajectory keeps this link explicit, and is the natural way to describe the star formation and chemical enrichment history at a given radius today: we solve the chemical evolution equations along the trajectories of the gas (the flow characteristics; Section~\ref{sec:general}). The resulting solutions are one-dimensional integrals that are fast enough to be embedded directly in a likelihood analysis, remaining easy to interpret because they connect the abundances we observe today to the past enrichment history of the gas. By predicting how the chemical abundances from prompt and delayed sources are distributed across the disc as a function of stellar birth radius and time, our formalism can be coupled to dynamical models of the Galaxy, which can follow how stars move in a given gravitational potential. Coupling the present solutions to a dynamical model of the MW might therefore be a natural way to build an efficient model to reconstruct the chrono-chemo-kinematic history of the MW disc that large spectroscopic surveys, combined with Gaia, have made it possible to measure.

This paper is organised as follows. In Section~\ref{sec:eqs} we present the governing equations of the model. In Section~\ref{sec:general} we derive the general integral solutions for an arbitrary radial velocity field using the method of characteristics. In Section~\ref{sec:data} we describe the observational data, and in Section~\ref{sec:model} the MW model and its free parameters. Section~\ref{sec:fitting} presents the fitting procedure, and Section~\ref{sec:results} the results. We summarise our conclusions in Section~\ref{sec:conclusions}.

\section{Governing equations}
\label{sec:eqs}
 
We assume cylindrical coordinates and the convention that radial gas flows directed towards the centre have positive velocities. So, following this convention, the assumed radial gas inflows in our models have velocities $v(r,t) > 0\,\text{km/s}$. The model only focuses on the abundance $Z$ of chemical elements like $\alpha$-elements and iron that are synthesised by core-collapse and Type Ia SNe. For core-collapse SNe we assume instantaneous recycling. The model properly takes into account the delay time distribution of Type Ia SNe. Finally, we assume that the gas in the galaxy disc is always well mixed at any time, $t$, and radius, $r$.
 
The model follows the evolution of the radial profiles of \textit{(i)} the gas surface mass density, $\Sigma_g(r,t)$, and \textit{(ii)} the surface mass density of a metal, $\Sigma_Z \equiv \Sigma_g Z(r,t)$, with abundance $Z$, by solving the following system of partial differential equations:

\begin{subequations}
\label{eq:system}
\begin{align}
\frac{\partial(\Sigma_g Z)}{\partial t}
- \frac{1}{r}\frac{\partial}{\partial r}\!\left(r\,v\,\Sigma_g Z\right)
&= -\hat{\epsilon}\,(1+\hat{\eta}-\hat{R})\,\Sigma_g Z \nonumber \\
&\quad + \langle y_Z\rangle_{\text{CC}}\,\hat{\epsilon}\,(1-\hat{R})\,\Sigma_g \nonumber \\
&\quad + \mathcal{S}_{\text{Ia}} + Z_{\rm inf}\,\hat{\mathcal{I}}, \label{eq:Z_system} \\[10pt]
\frac{\partial \Sigma_g}{\partial t}
- \frac{1}{r}\frac{\partial}{\partial r}\!\left(r\,v\,\Sigma_g\right)
&= -\hat{\epsilon}\,(1+\hat{\eta}-\hat{R})\,\Sigma_g + \hat{\mathcal{I}}, \label{eq:gas}
\end{align}
\end{subequations}

where the various terms and quantities can be summarised as follows:
\begin{enumerate}
    \item $\psi = \hat{\epsilon}\,\Sigma_g$ is the star-formation rate (SFR) surface density, which follows a linear Schmidt-Kennicutt relation \citep{Kennicutt1998} with a star formation efficiency $\hat{\epsilon}(r,t)$; 
    \item $\hat{\eta}(r,t)$ is the mass-loading factor that regulates the intensity of the outflow rate $\hat{\mathcal{O}}=\hat{\eta}\,\psi$; 
    \item $\hat{R}$ is the return mass fraction, treated as constant; 
    \item $\langle y_Z\rangle_{\text{CC}}$ is the average stellar nucleosynthetic yield from core-collapse SNe per unit stellar mass formed, also treated as constant; 
    \item $\hat{\mathcal{I}}(r,t)$ is the analytic gas infall rate; 
    \item $\mathcal{S}_{\text{Ia}}(r,t)$ is the Type~Ia SN enrichment rate; 
    \item $Z_{\rm inf}(r,t)$ is the abundance of the relevant element in the infalling gas, so that $Z_{\rm inf}=0$ simulates pristine (primordial) accretion while $Z_{\rm inf}>0$ describes pre-enriched infall.
\end{enumerate}
 
\textit{Type Ia supernovae.} SNe~Ia are exploding white dwarfs in a binary system that release chemical elements in the ISM on a distribution of delay times from the formation of the binary system. In our Lagrangian framework, the Type~Ia SN enrichment is followed along the gas trajectory. The relevant star formation history is the one corresponding to the position of the parcel along its path, $\psi(\xi,\tau)=\hat{\epsilon}(\xi,\tau)\,\Sigma_g(\xi,\tau)$, in which the star formation efficiency and gas surface density are evaluated at the radius occupied by the parcel at each time. The Type~Ia SN rate is then obtained by convolving this star formation history with our assumed delay-time distribution DTD $\propto t^{-1.1}$ from \citet{Maoz2017}, along the characteristic:

\begin{equation}
\mathcal{S}_{\text{Ia}}(\xi,\tau) = \langle y_Z \rangle_{\text{Ia}}
\int_{\tau_{\text{Ia}}}^{\tau} \text{DTD}_{\text{Ia}}(s)\,
\psi(\xi,\tau-s)\,\mathrm{d}s,
\label{eq:Ia_source}
\end{equation}

where 
\begin{enumerate}
    \item $\langle y_{Z}\rangle_{\text{Ia}}$ is the average mass of a specific chemical element ejected by each SN~Ia event;
    \item  $\tau_{\text{Ia}}=0.15\,\text{Gyr}$ is the assumed minimum delay time for the explosion of a Type Ia SN from the formation of the binary system \citep{Johnson2021,Vincenzo2021,Weinberg2024};
    \item the DTD is normalised so as to have $2$ SNe~Ia per $10^{3}\,\text{M}_{\odot}$ of stellar mass formed \citep{Bell2003,Maoz2014,Vincenzo2017,Vincenzo2021}.
\end{enumerate} 
Since the mass ejected by SNe Ia is statistically negligible in the overall simulation volume, the gas equation~(\ref{eq:gas}) is unaffected by $\mathcal{S}_{\text{Ia}}$.

\section{General solution for an arbitrary velocity field}
\label{sec:char}
\label{sec:general}
 
We solve the system of equations~(\ref{eq:system}) for a generic velocity $v(r,t)$, following a Lagrangian approach, by using the method of characteristics. A characteristic represents the trajectory of an individual parcel of gas as it flows radially inward, providing mathematically a transformation of coordinates to study radial gas flows in a co-moving frame. 

Consider all gas that -- at the time $t$ -- resides at a given radius, $r$. For an inward gas flow, the characteristic can be used to parameterise all positions where the gas resided at previous times. In this co-moving frame, the spatial advection terms in the system of equations~(\ref{eq:system}) vanish, reducing the partial differential equations to ordinary differential equations in time. We label each characteristic by its initial radius $\xi$ and, for notational convenience, we write $f(\xi,\tau)\equiv f(r(\xi,\tau),\tau)$. The characteristic curves are

\begin{subequations}
\label{eq:char_curve}
\begin{align}
&\frac{\mathrm{d}r}{\mathrm{d}\tau} = -\,v(r,\tau),
\\[5pt]
&r(\xi,0) =\xi,
\end{align}
\end{subequations}

where $\mathrm{d}r/\mathrm{d}\tau= -v(r,\tau)<0$ traces the inward motion. These curves are common to both equations of system~(\ref{eq:system}), because they both share the same advection operator $\partial_t - v\,\partial_r$. For a prescribed $v$, the characteristics only depend on the assumed velocity field $v(r,t)$, determining the evolution of $\Sigma_g$ and $Z$.
 
\subsection{Gas surface density}
\label{sec:gas_char}
 
By expanding the divergence on the left-hand side of equation~(\ref{eq:gas}) with the product rule, $\frac{1}{r}\partial_r(r v\Sigma_g) = (\partial_r v + v/r)\,\Sigma_g + v\,\partial_r\Sigma_g$, the time-derivative and the $v\,\partial_r\Sigma_g$ term combine into $\partial_t\Sigma_g - v\,\partial_r\Sigma_g$, which is the Lagrangian derivative $\mathrm{d}\Sigma_g/\mathrm{d}\tau$ along a characteristic curve (equation~\ref{eq:char_curve}). The remaining term $(\partial_r v + v/r)\,\Sigma_g$ is moved to the right-hand side, and the gas equation becomes a linear ordinary differential equation along each characteristic:

\begin{equation}
\frac{\mathrm{d}\Sigma_g}{\mathrm{d}\tau}\bigg|_\xi
= -\Big[\,\hat{\epsilon}\,(1+\hat{\eta}-\hat{R})
- \big(\partial_r v + v/r\big)\,\Big]\,\Sigma_g + \hat{\mathcal{I}}.
\label{eq:Sigma_char}
\end{equation}

\begin{enumerate} 
\item The term $(\partial_r v + v/r)$ in equation~(\ref{eq:Sigma_char}) accounts for two distinct effects, which both increase $\Sigma_{g}$: \textit{(i)} the $v/r$ component represents cylindrical convergence, whereby an inward flow is squeezed into increasingly smaller annular areas ($2\pi r\,\mathrm{d}r$); \textit{(ii)} the $\partial_r v$ component instead characterises how a velocity gradient affects the gas surface mass density at a given radius: a positive spatial gradient ($\partial_r v > 0$) causes the faster inflowing gas from the outer regions to pile up against the slower gas ahead of it, further amplifying $\Sigma_g$.
\item The infall rate $\hat{\mathcal{I}}$ in equation~(\ref{eq:Sigma_char}) is the only source of gas at a given radius and time, and increases $\Sigma_{g}$.
\item The remaining term $\hat{\epsilon}\,(1+\hat{\eta}-\hat{R})\,\Sigma_g$ represents the net gas removal, combining consumption by star formation ($\hat{\epsilon}$) and loss through galactic outflows ($\hat{\eta}\,\hat{\epsilon}$, assumed proportional to the star formation rate), partly offset by the return of gas from dying stars (through $\hat{R}$).
\end{enumerate}

\noindent With the
integrating factor

\begin{equation}
\mu_g(\xi,\tau)=\exp\!\,\left(\int_0^{\tau}
\Big[\hat{\epsilon}\,(1+\hat{\eta}-\hat{R})
- \big(\partial_r v + v/r\big)\Big]\,\mathrm{d}\tau'\right),
\label{eq:mu_g}
\end{equation}

the solution of equation~(\ref{eq:Sigma_char}) is

\begin{equation}
\Sigma_g(\xi,\tau)=\frac{1}{\mu_g(\xi,\tau)}\!\left[\,\Sigma_{g,0}(\xi)
+\int_0^{\tau}\hat{\mathcal{I}}(\xi,\tau')\,\mu_g(\xi,\tau')\,
\mathrm{d}\tau'\right],
\label{eq:Sigma_sol}
\end{equation}

where $\Sigma_{g,0}(\xi)=\Sigma_g(\xi,0)$ is the initial gas surface mass density along the characteristic labelled by $\xi$, i.e.\ the density of the gas that resides at radius $\xi$ at time $t=0$ and reaches radius $r=\xi-v\tau$ at time $t=\tau$.

By looking at equation~(\ref{eq:Sigma_sol}), as the gas flows inwards, the main driver for the increase of its density $\Sigma_{g}$ over time is given by gas accretion, which is quantified by the integral in equation~(\ref{eq:Sigma_sol}). The integrating factor $\mu_g$ at the denominator of equation~(\ref{eq:Sigma_sol}) encodes the net balance between gas consumption (through star formation and outflows) and the density amplification due to the flow geometry. The larger the consumption, the larger $\mu_g$ and hence the lower the resulting $\Sigma_{g}$; conversely, the larger the geometric amplification, the smaller $\mu_g$ and the larger $\Sigma_{g}$.
 
\subsection{Metallicity}
\label{sec:Z}
 
Following a procedure analogous to that used for the gas surface mass density [equation~(\ref{eq:Sigma_char})], we expand the divergence in equation~(\ref{eq:Z_system}) and substitute $\partial_t\Sigma_g-\frac{1}{r}\partial_r(r\Sigma_g v)$ using the gas equation~(\ref{eq:gas}); interestingly, in doing so, the loss terms $-\hat{\epsilon}(1+\hat{\eta}-\hat{R})\Sigma_g Z$ cancel. Dividing by $\Sigma_g$, we obtain an equation for the metallicity $Z$ alone:

\begin{equation}
\frac{\partial Z}{\partial t} - v\,\frac{\partial Z}{\partial r}
= \hat{\epsilon}\,(1-\hat{R})\,\langle y_Z\rangle_{\text{CC}}
+ \frac{\mathcal{S}_{\text{Ia}}}{\Sigma_g}
+ \frac{\hat{\mathcal{I}}}{\Sigma_g}\left(Z_{\rm inf}-Z\right).
\label{eq:Z}
\end{equation}

It is interesting to note that the term $\Sigma_g(\partial_r v + v/r)$ in equation~(\ref{eq:Z_system}) cancels exactly in the passage from $\Sigma_Z$ to $Z$, leaving the advective operator $\partial_t Z - v\,\partial_r Z$ with the same mathematical expression as in the Cartesian case. On the right-hand side of equation~(\ref{eq:Z}), the sources are core-collapse SNe,  SNe~Ia, and the net effect of infall. Pristine accretion dilutes $Z$ at a rate $-(\hat{\mathcal{I}}/\Sigma_g)\,Z$, while an enriched infall with metal abundance $Z_{\rm inf}$ adds metals at a rate $+Z_{\rm inf}\hat{\mathcal{I}}/\Sigma_g$; in the absence of the other source terms, the two infall terms would drive $Z$ towards the infall abundance $Z_{\rm inf}$.
 
Along the same characteristics that are used to calculate $\Sigma_g$ (see equation~\ref{eq:char_curve}), equation~(\ref{eq:Z}) takes the form of the following linear ordinary differential equation:

\begin{equation}
\frac{\mathrm{d}Z}{\mathrm{d}\tau}\bigg|_\xi
= \hat{\epsilon}\,(1-\hat{R})\,\langle y_Z\rangle_{\text{CC}}
- \frac{\hat{\mathcal{I}}}{\Sigma_g}\,Z
+ \frac{\mathcal{S}_{\text{Ia}}}{\Sigma_g}
+ \frac{Z_{\rm inf}\,\hat{\mathcal{I}}}{\Sigma_g},
\label{eq:Z_char}
\end{equation}

with $\Sigma_g$ given by equation~(\ref{eq:Sigma_sol}).
 
Defining the integrating factor as follows:

\begin{equation}
\mu_Z(\xi,\tau)=\exp\!\left(\int_0^{\tau}
\frac{\hat{\mathcal{I}}(\xi,\tau')}{\Sigma_g(\xi,\tau')}\,\mathrm{d}\tau'\right),
\label{eq:mu_Z}
\end{equation}

the integral solution is

\begin{align}
Z(\xi,\tau) &= \frac{1}{\mu_Z(\xi,\tau)}\Bigg[\,Z_0(\xi) \nonumber \\
&\quad +\int_0^{\tau}\!\left(\hat{\epsilon}\,(1-\hat{R})\,\langle y_Z\rangle_{\text{CC}}
+\frac{\mathcal{S}_{\text{Ia}}}{\Sigma_g}
+\frac{Z_{\rm inf}\,\hat{\mathcal{I}}}{\Sigma_g}\right)\!\,\mu_Z(\xi,\tau')\,
\mathrm{d}\tau'\Bigg],
\label{eq:Z_sol}
\end{align}

By looking at equations~(\ref{eq:mu_Z})-(\ref{eq:Z_sol}), as the gas flows inwards, the larger the accreted gas mass $\hat{\mathcal{I}}$ relative to the gas already present $\Sigma_{g}$, the larger $\mu_Z$, and the lower the first term in parentheses in equation~(\ref{eq:Z_sol}) with the contribution of the initial metallicity $Z_0$, reducing its effect on the final value $Z$. A second effect of infall, less evident from equation~(\ref{eq:Z_sol}), is its attenuation of the metallicity enrichment by core-collapse and Type~Ia SNe. This second effect becomes more evident if the prefactor $1/\mu_Z$ in equation~(\ref{eq:Z_sol}) is brought inside the integral, namely

\begin{align}
Z(\xi,\tau) ={}& \frac{Z_0(\xi)}{\mu_Z(\xi,\tau)} \nonumber \\
&+ \int_0^{\tau}\!\left(\hat{\epsilon}\,(1-\hat{R})\,\langle y_Z\rangle_{\text{CC}}
+ \frac{\mathcal{S}_{\text{Ia}}}{\Sigma_g}
+ \frac{Z_{\rm inf}\,\hat{\mathcal{I}}}{\Sigma_g}\right)
\frac{\mu_Z(\xi,\tau')}{\mu_Z(\xi,\tau)}\,\mathrm{d}\tau' .
\label{eq:Z_sol_split}
\end{align}

The source term in parentheses within the integral of equation~(\ref{eq:Z_sol_split}) is weighted by the ratio

\begin{equation}
\frac{\mu_Z(\xi,\tau')}{\mu_Z(\xi,\tau)}
= \exp\!\left(-\int_{\tau'}^{\tau}\frac{\hat{\mathcal{I}}}{\Sigma_g}\,
\mathrm{d}\tau''\right) \le 1,
\label{eq:mu_ratio}
\end{equation}

meaning that the metals released by core-collapse and Type~Ia SNe at time $\tau'$ are diluted by the gas that falls in from $\tau'$ up to the present time $\tau$. The earlier the metals are produced, the larger the dilution, as they have been exposed to the entire subsequent infall; those produced late are diluted the least.
 
We note that, when $v=0$, the characteristics become vertical lines, $\xi$ coincides with the Eulerian radius $r$, and equations~(\ref{eq:Sigma_sol}) and~(\ref{eq:Z_sol}) reduce to independent one-zone models at each radius.
 
\subsection{Computational sequence}
 
The infall rate $\hat{\mathcal{I}}(r,\tau)$ is assumed to be an analytic function, and $\mathcal{S}_{\text{Ia}}$ depends only on the past history of $\Sigma_g$ (which is itself independent of $Z$ in our model). Therefore, the two integral solutions can be computed in sequence: we first calculate $\Sigma_g$ and then the evolution of $Z$. In particular, for a given characteristic $\xi$ and target time $\bar{\tau}$, the computation proceeds as follows:
\begin{enumerate}
\item \textit{Characteristic.} We integrate $\mathrm{d}r/\mathrm{d}\tau = -v(r,\tau)$ with $r(\xi,0) = \xi$ to obtain the trajectory $r(\xi,\tau)$ for all $\tau \in [0,\bar{\tau}]$.
\item \textit{Integrating factor $\mu_g$.} We calculate the integrating factor for the gas surface density, equation~(\ref{eq:mu_g}), numerically; this involves the quantity $\partial_r v + v/r$, calculated along $r(\xi,\tau')$.
\item \textit{Gas surface density $\Sigma_g$.} We calculate the gas surface mass density from equation~(\ref{eq:Sigma_sol}) numerically, which also provides the star formation rate $\psi(\xi,\tau') = \hat{\epsilon}(\xi,\tau')\, \Sigma_g(\xi,\tau')$ for all $\tau' \in [0,\bar{\tau}]$.
\item \textit{SN~Ia convolution.} We calculate the Type~Ia SN rate, equation~(\ref{eq:Ia_source}), numerically along the characteristic, namely

\begin{equation}
\mathcal{S}_{\text{Ia}}(\xi,\tau) = \langle y_Z \rangle_{\text{Ia}}
\int_{\tau_{\text{Ia}}}^{\tau} \text{DTD}_{\text{Ia}}(s)\,
\psi(\xi,\tau-s)\,\mathrm{d}s
\label{eq:Ia_char}
\end{equation}

for each $\tau \in [0,\bar{\tau}]$, using the star formation rate $\psi(\xi,\tau')$ from step~\textit{(iii)}.
\item \textit{Integrating factor $\mu_Z$.} We calculate the integrating factor for the metallicity, equation~(\ref{eq:mu_Z}), numerically, using the gas surface mass density $\Sigma_g(\xi,\tau')$ from step~\textit{(iii)}.
\item \textit{Metallicity $Z$.} We calculate the metallicity from equation~(\ref{eq:Z_sol}) numerically, using the Type~Ia SN rate $\mathcal{S}_{\text{Ia}}$ from step~\textit{(iv)} and the integrating factor $\mu_Z$ from step~\textit{(v)}.
\end{enumerate}
Each step requires the calculation of a single numerical integral over $[0,\bar{\tau}]$ and depends only on quantities computed in the previous steps.
 
\subsection{Recovery of the Eulerian fields}
 
The integral solutions~(\ref{eq:Sigma_sol})--(\ref{eq:Z_sol}) are expressed in Lagrangian coordinates $(\xi,\tau)$. To recover the Eulerian fields $\Sigma_g(r,t)$ and $Z(r,t)$ on a fixed spatial grid $\{r_j\}$ at a given time $\bar{\tau}$, we retrieve for each grid point $r_j$ the Lagrangian label $\bar{\xi}$ of the characteristic passing through $(r_j,\bar{\tau})$, defined by

\begin{equation}
r(\bar{\xi},\,\bar{\tau}) = r_j .
\end{equation}

This equation is inverted to find $\bar{\xi}(r_j,\bar{\tau})$, either analytically when the characteristic map admits a closed form, or numerically via root-finding. The integral solutions for the metallicity $Z$ and the gas surface density $\Sigma_{g}$ as a function of time are then determined along the characteristic $\bar{\xi}$ corresponding to $(r_j,\bar{\tau})$, yielding $\Sigma_g(r_j,\bar{\tau})$ and $Z(r_j,\bar{\tau})$.

\section{Milky Way observational data}
\label{sec:data}
 
The model predictions are compared with the chemical abundances as measured in a sample of MW red giant stars at different Galactocentric distances and Galactic latitudes by the Apache Point Observatory Galactic Evolution Experiment (APOGEE, \citealt{Majewski2017}) of the Sloan Digital Sky Survey IV (SDSS IV, \citealt{Blanton2017}); in particular, we make use of SDSS IV APOGEE~Data Release 17 \citep[][]{Abdurrouf2022}. The position of the stars in cylindrical coordinates alongside an estimate of their ages are taken from the value-added \texttt{astroNN} catalogue \citep{Leung2019,Mackereth2019}.
 
We apply the following quality cuts, broadly following \citet{Weinberg2022}:
\begin{enumerate}
    \item \texttt{ASPCAPFLAG}$\,=0$, to only include stars without problematic spectra or unreliable stellar parameters flagged by the APOGEE Stellar Parameters and Abundances Pipeline (ASPCAP; see also \citealt{garciaperez2016,Holtzman2018,jonsson2020});
    \item \texttt{EXTRATARG}$\,=0$, to only include stars in the main APOGEE survey;
    \item signal-to-noise ratio $\mathrm{S/N}>100$;
    \item surface gravity $1.0\le\log \left(\, g/[\mathrm{cm}\,\mathrm{s}^{-2}]\,\right)\le2.5$;
    \item effective temperature $4000\le T_{\rm eff}/[\mathrm{K}]\le4600$; 
\end{enumerate}
The selection window in surface gravity and effective temperature is made to remove red clump stars from the sample (see \citealt{Vincenzo2021,Weinberg2022} for more details). 
After applying these quality cuts, our final sample consists of $50\,288$ stars. We adopt oxygen as the $\alpha$-element tracer because of its prominence among the $\alpha$-elements and work in the [O/Fe]--[Fe/H] plane. We note that the chemical evolution of O can be affected by cross-pollution in binaries (e.g., see \citealt{deMink2009,Pepe2025,Pepe2026}). The APOGEE sample is then cross-matched with the \texttt{astroNN} catalogue by \texttt{APOGEE\_ID}.
 
The disc is divided into Galactocentric annuli of width $\Delta R = 1$~kpc, and at each radius we restrict the observational sample to stars close to the midplane, $|z|\le 0.5$~kpc. Throughout we adopt the solar reference abundances of \citet{Asplund2009}, corresponding to mass fractions $(Z_{\rm Fe})_\odot = 13.7\times10^{-4}$ and $(Z_{\rm O})_\odot = 73.3\times10^{-4}$ \citep{Weinberg2022}.
 
The predictions of our models for the radial chemical abundance gradients of [Fe/H] and [O/H] as a function of time are compared with the observations of the SDSS Open Cluster Chemical Abundances and Mapping (OCCAM) survey \citep{Myers2022} in a sample of open clusters, based on APOGEE~DR17. We adopt the OCCAM high-quality sample and use the Galactocentric radii, $R_{\rm GC}$, ages, and mean cluster abundances $\mathrm{[Fe/H]}$ and $\mathrm{[O/Fe]}$ provided in their catalogue. The clusters are divided into the same four age bins used by \citet{Myers2022}.

\section{Milky Way model and free parameters}
\label{sec:model}
 
\subsection{Gas accretion}
\label{sec:model-gas-accretion}
Because we aim to fit the observed bimodality between thick- and thin-disc stars in the [O/Fe]--[Fe/H] abundance distribution across the disc, we assume a two-infall model for the chemical evolution of the MW disc \citep{Chiappini1997,Spitoni2019,Palla2020,Spitoni2021}. According to this model, the disc is assembled following two exponentially declining accretion episodes: a first one on a short timescale, giving rise to the high-[$\alpha$/Fe] thick-disc stellar populations, and a second, more prolonged one, giving rise to the low-[$\alpha$/Fe] thin-disc populations. To reproduce the chemical abundance gradients, the infall timescales increase, on average, when moving outwards, developing an inside-out growth of the Galaxy disc as a function of time \citep{Matteucci1989,Johnson2021}.
Following the formalism outlined in \citet{Spitoni2019,Spitoni2020,Spitoni2021}, we model the accretion
rate as the sum of two exponential terms as follows:

\begin{equation}
\begin{split}
\hat{\mathcal{I}}(r,t) =\ &\frac{\Sigma_1(r)}{\tau_1(r)\left[\,1-e^{-t_G/\tau_1(r)}\,\right]}\,e^{-t/\tau_1(r)} \\
+\ &\frac{\Sigma_2(r)}{\tau_2(r)\left[\,1-e^{-(t_G-t_{\rm max})/\tau_2(r)}\,\right]}\,\Theta(t-t_{\rm max})\,e^{-(t-t_{\rm max})/\tau_2(r)},
\end{split}
\label{eq:two_infall}
\end{equation}

where $\tau_1(r)$ and $\tau_2(r)$ are the thick- and thin-disc infall timescales, respectively, $t_{\rm max}$ is the time marking the onset of the second accretion episode, and $\Theta$ is the Heaviside function. At each radius $r$, the time integrals of the two infall episodes are normalised to give the  cumulative accreted surface mass densities, $\Sigma_1(r)$ for the thick disc and $\Sigma_2(r)$ for the thin disc. The total accreted  surface mass density is assumed to follow an exponential radial profile \citep{Lacey1985,Matteucci1989,Spitoni2017,Palla2020,Spitoni2021}:

\begin{equation}
\Big[\Sigma_1(r) + \Sigma_2(r) \Big]\propto e^{-(r-R_\odot)/r_d},
\label{eq:sigma_profile}
\end{equation}

where $R_\odot = 8$~kpc is the Solar radius and $r_d = 3.5$~kpc is the assumed scale length.
 
Following a similar procedure like in \citet{Spitoni2021}, the ratio $\Sigma_2(r)/\Sigma_1(r)$ is a free parameter of the model, determining the relative amount of gas accreted in the two episodes. In terms of this quantity, the accretion rate in equation~(\ref{eq:two_infall}) can be rewritten as

\begin{equation}
\begin{split}
\hat{\mathcal{I}}(r,t) = \mathcal{A}\,e^{-(r-R_\odot)/r_d}\bigg[\,&
\frac{1}{1+\Sigma_2/\Sigma_1}\,
\frac{e^{-t/\tau_1}}{\tau_1\left(1-e^{-t_G/\tau_1}\right)} \\
+\ &\frac{\Sigma_2/\Sigma_1}{1+\Sigma_2/\Sigma_1}\,
\frac{\Theta(t-t_{\rm max})\,e^{-(t-t_{\rm max})/\tau_2}}
{\tau_2\left(1-e^{-(t_G-t_{\rm max})/\tau_2}\right)}\bigg],
\end{split}
\label{eq:infall_ratio}
\end{equation}

where $\mathcal{A}$ is calculated in our model so as to match the present-day total stellar mass of the Milky Way. In particular, the model iteratively rescales the infall amplitude and integrates the resulting stellar surface-density profile $\Sigma_\star(r,t_G)$ over the disc until the total stellar mass reaches $M_\star(t_G) = (5.17\pm 1.11)\times10^{10}\,M_\odot$ \citep{LicquiaNewman2015}, which is also adopted as reference by the MW chemical evolution models of \citet{Johnson2021}. This calculation is performed using the following equation:

\begin{equation}
\begin{split}
M_\star(t_G) &= \int \Sigma_\star(r,t_G)\,2\pi r\,\mathrm{d}r \\
&= (1-\hat{R})\int_0^{t_G}\!\!\int \psi(r,t)\,2\pi r\,\mathrm{d}r\,\mathrm{d}t .
\end{split}
\label{eq:Mstar_calib}
\end{equation}

Since $M_\star$ is approximately proportional to $\mathcal{A}$, convergence is typically reached in $2$ iterations.
 
In the absence of radial flows ($v=0\,\mathrm{km\,s^{-1}}$), the $M_\star$ calibration above leaves the abundance tracks unchanged because each annulus evolves independently, and the predicted chemical abundances do not change for different infall normalisations. When radial gas flows are included, the gas reaching a given radius is influenced by the chemical evolution of the outer regions it originated from, so that the abundances become sensitive to the surface-density profile across the disc, including outer regions where the data may be too sparse to constrain the chemical-abundance fit. To keep the fitting procedure tractable, we first determine the best-fit model parameters by setting $v=0\,\mathrm{km\,s^{-1}}$ (Section~\ref{sec:fitting}); we then keep these best-fit parameters fixed and use the general $v\neq0$ solution to test the effect of radial gas flows on the chemical evolution (Section~\ref{sec:results_stars}). For each value of $v$, the infall is rescaled so as to reproduce the observed present-day stellar mass of the Milky Way, $M_\star(t_G)=(5.17\pm1.11)\times10^{10}\,M_\odot$.
 
\subsection{Timescales}
 
The thick-disc phase takes place during the time interval $[0, t_{\rm max}]$; for short $\tau_1$, the first infall episode may exhaust itself well before $t_{\rm max}$, leaving a quiescent gap between the two star-forming phases. The thin-disc phase then is assumed to run for $10\,\text{Gyr}$ from $t_{\rm max}$ to $t_{\rm max} + 10$~Gyr. The present-day age of the Galaxy is therefore $t_G = 10\,\mathrm{Gyr} + t_{\rm max}$, and the age of a star formed at time $t$ is $(10 + t_{\rm max}) - t$.
 
The duration of the thick-disc phase (including a possible quiescent gap) plus that of the thin-disc phase cannot exceed the age of the Galaxy, namely $t_{\rm max} + 10 \leq t_G$; this gives the constraint $t_{\rm max} \leq 3.8$~Gyr. In our model, to minimise the number of free parameters, we fix $t_{\rm max} = 3.8$~Gyr, which is consistent with the findings of previous works for the thick-to-thin disc transition epoch (e.g., see \citealt{Palla2020,Spitoni2021}, both assuming a two-infall model).
 
\subsection{Constraining the thin-disc infall timescale from stellar ages}
 
We empirically estimate how the thin-disc infall timescale, $\tau_2(r)$, changes as a function of radius $r$ from the analysis of the \texttt{astroNN} stellar age distribution at different Galactocentric distances \citep{Leung2019,Mackereth2019}. The measured ages are obtained by cross-matching the \texttt{astroNN} catalogue with our reference APOGEE-DR17 red giant sample, using the quality cuts described in Section~\ref{sec:data}. This way the age and abundance analyses are performed on the same consistent sample of stars.
 
In our calculation, we consider concentric rings with width $\Delta R = 1.0\,\text{kpc}$, centred at radii $R = 4$-$11\,\text{kpc}$ and vertical heights $|z|\le 0.5\,\text{kpc}$. In each ring, we separate thin- and thick-disc stars using the dividing line in [O/Fe]--[Fe/H] proposed by \citet{Weinberg2022}:

\begin{equation}
[\text{O/Fe}]_{\rm boundary} =
\begin{cases}
0.12 - 0.13\,[\text{Fe/H}], & [\text{Fe/H}] < 0,\\
0.12, & [\text{Fe/H}] \ge 0,
\end{cases}
\label{eq:boundary}
\end{equation}

with stars above (below) this boundary assigned to the high-[$\alpha$/Fe] thick-disc (low-[$\alpha$/Fe] thin-disc) stellar populations. Then, in each annulus centred at $r$, $\tau_2$ is estimated from the mean stellar age of the thin-disc population as

\begin{equation}
\tau_2(r) = 10\,\text{Gyr} - \langle\, t_{\rm age}(r)\,\rangle_{\rm thin-disc}.
\label{eq:tau2}
\end{equation}

assuming that $10\,\text{Gyr}$ is the duration of the thin-disc phase. The resulting values of $\tau_2(r)$, listed in Table~\ref{tab:tau_thin}, are interpolated to evaluate the timescale at any radius.

We note that the line defined by equation~(\ref{eq:boundary}) is used to separate thin- and thick-disc stars in the observational sample to calculate the observed age distribution of thin-disc stars, while thick-disc and thin-disc stellar populations in the model are separated according to their predicted formation time (stars formed before or after $t_{\text{max}}$, respectively). Because the two-infall model predicts that the [O/Fe]--[Fe/H] abundance pattern crosses the line defined by equation~(\ref{eq:boundary}) more than once, we caution that, if we identified thick- and thin-disc stars in the model only through equation~(\ref{eq:boundary}), there would be some contamination between the two populations; in particular, there are thin-disc stars in the two-infall model (i.e., stars born after $t_{\text{max}}$) with abundance patterns above that line, and they would then be classified as thick-disc stars based solely on equation~(\ref{eq:boundary}). Similarly, there are thick-disc stars in the two-infall model (i.e., stars born before $t_{\text{max}}$) that have abundance patterns below that line and would be classified as thin-disc stars.

\begin{table}
\centering
\caption{Thin-disc infall timescale $\tau_2$ as a function of Galactocentric radius.}
\label{tab:tau_thin}
\begin{tabular}{cc}
\hline\hline
$R$ (kpc) & $\tau_2$ (Gyr) \\
\hline
4.0  & 4.59 \\
5.0  & 4.79 \\
6.0  & 4.79 \\
7.0  & 4.78 \\
8.0  & 4.96 \\
9.0  & 5.16 \\
10.0 & 5.36 \\
11.0 & 5.59 \\
\hline
\end{tabular}
\tablefoot{Derived from the mean stellar age of the low-$\alpha$ population in each annulus (Section~\ref{sec:model}).}
\end{table}
 
\subsection{Star formation and nucleosynthesis}
\label{sec:starformation}
The SFR is assumed to follow a linear Schmidt-Kennicutt law $\psi=\hat{\epsilon}\,\Sigma_g$ \citep{Kennicutt1998}. The star formation efficiency differs between the two infall episodes as follows:

\begin{equation}
\hat{\epsilon}(r,t) =
\begin{cases}
\hat{\epsilon}_{\rm thick}(r), & t<t_{\rm max},\\[2pt]
\hat{\epsilon}_{\rm thin}(r),  & t\ge t_{\rm max},
\end{cases}
\end{equation}

with the switch between the thick disc star formation efficiency, $\hat{\epsilon}_{\rm thick}$, and the thin disc star formation efficiency, $\hat{\epsilon}_{\rm thin}$, happening at the time $t = t_{\rm max}$, corresponding to the onset of the thin-disc infall episode.
 
The nucleosynthetic yields are held constant, assuming the following values for core-collapse and Type~Ia SNe:
\begin{enumerate}
    \item $\langle y_{\rm O}\rangle_{\rm CC}=1.022\times10^{-2}$ and
$\langle y_{\rm Fe}\rangle_{\rm CC}=5.6\times10^{-4}$ correspond to the average yields of oxygen and iron from core-collapse SNe per unit stellar mass formed assuming a \citet{Kroupa1993} initial mass function \citep{Vincenzo2016,Vincenzo2017};
\item $\langle y_{\rm O}\rangle_{\rm Ia}=1.43\times10^{-1}\,\mathrm{M}_\odot$ and $\langle y_{\rm Fe}\rangle_{\rm Ia}=0.77\,\mathrm{M}_\odot$ correspond to the yields of oxygen and iron for each Type Ia SN event from \citet{Iwamoto1999};
\item $\hat{R}=0.285$ is the return mass fraction assuming a \citet{Kroupa1993} initial mass function \citep{Vincenzo2016,Vincenzo2017}.
\end{enumerate}
 
\subsection{Boundary conditions}
In our model the infall term in equation~(\ref{eq:gas}) represents the only source of gas. After being accreted, such gas is then transported inward by radial gas flows, assuming a certain velocity field. We impose a maximum radius for the gaseous disc  $R_{\rm out}=20$~kpc, beyond which no gas is accreted and no stars form, in agreement with the boundary conditions adopted in the multi-zone models of \citet{Mott2013} for the Milky Way and \citet{Spitoni2013} for M31. Gas located within $R_{\rm out}$ flows inward, feeding the inner disc, but no characteristic is crossing the boundary bringing material from larger radii. Because infall is the only gas source, this boundary condition is implemented through two conditions, one on the accretion term and another on the star formation efficiency, as follows:

\begin{equation}
\hat{\mathcal{I}}(r,t) = 0 \quad\text{and}\quad \hat{\epsilon}(r,t) = 0
\qquad \text{for } r > R_{\rm out},
\label{eq:Rout}
\end{equation}

Since no gas is gained beyond $R_{\rm out}$ (the infall rate is set to zero), the amount of gas that can be transported across that radius from the outside is always zero by construction. Characteristics originating at $\xi > R_{\rm out}$ therefore start essentially empty of gas and remain chemically inert until, flowing inward, they cross $R_{\rm out}$; only then do they begin to accumulate gas and form stars. In this way the outer boundary acts as a donor of gas to the inner disc, exactly as intended.

\begin{table}
\centering
\caption{Best-fit parameters of the two-infall model as a function of Galactocentric radius.}
\label{tab:bestfit}
\begin{tabular}{cccccc}
\hline\hline
$R$ & $\tau_2$ & $\tau_1$ & $\Sigma_2/\Sigma_1$ & $\hat{\epsilon}_{\rm thin}$ & $\hat{\epsilon}_{\rm thick}$ \\ 
(kpc) & (Gyr) & (Gyr) & & (Gyr$^{-1}$) & (Gyr$^{-1}$) \\
\hline
4.0  & 4.59 & 0.20 & 7.2  & 2.06 & 1.50 \\
5.0  & 4.79 & 0.32 & 8.9  & 1.66 & 1.70 \\
6.0  & 4.79 & 0.33 & 10.3 & 1.36 & 1.57 \\
7.0  & 4.78 & 0.30 & 10.1 & 1.17 & 1.54 \\
8.0  & 4.96 & 0.38 & 13.0 & 0.87 & 1.41 \\
9.0  & 5.16 & 0.41 & 20.2 & 0.69 & 1.44 \\
10.0 & 5.36 & 0.62 & 21.6 & 0.58 & 1.43 \\
11.0 & 5.59 & 0.56 & 24.8 & 0.50 & 1.16 \\
\hline
\end{tabular}
\tablefoot{The thin-disc infall timescale $\tau_2$ is fixed from stellar ages (Table~\ref{tab:tau_thin}); $t_{\rm max}=3.8$~Gyr and $\hat{\eta}=0$ are fixed at all radii.}
\end{table}

 \begin{figure*}
    \centering
    \includegraphics[width=0.8\textwidth]{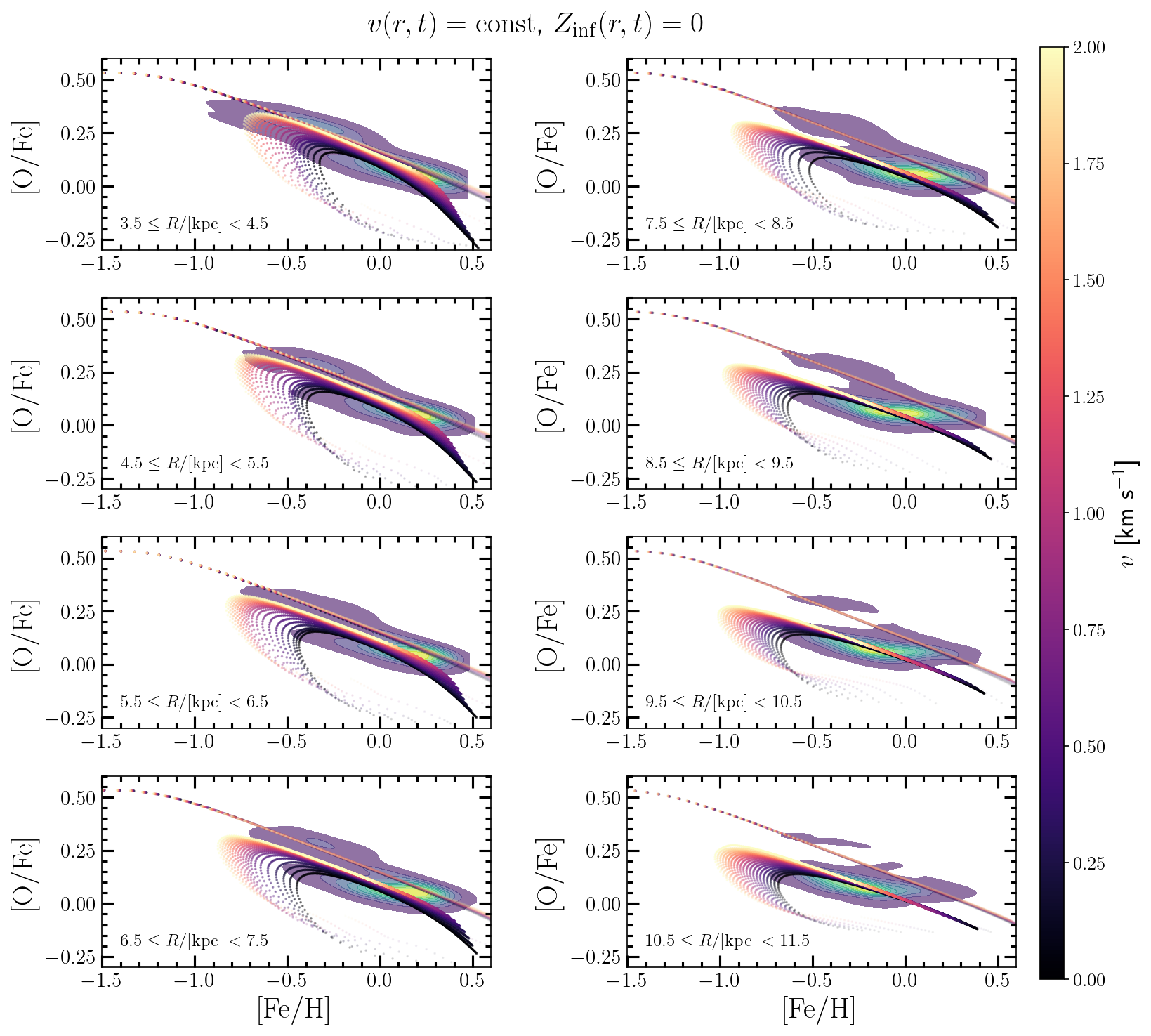}
    \caption{The $[\mathrm{O/Fe}]$--$[\mathrm{Fe/H}]$ abundance plane in eight radial bins spanning $3.5 \leq R/\mathrm{kpc} < 11.5$. The background colour map shows the stellar number density of APOGEE DR17 stars estimated via a 2D Gaussian kernel density estimator, with regions below 2\% of the peak density shown in white. Stars are selected within $|z| \leq 0.5\,\mathrm{kpc}$ of the Galactic plane. The coloured tracks show the predictions of the chemical evolution model for constant radial inflow velocities $v=0$--$2\,\mathrm{km\,s^{-1}}$ (colour bar on the right), evaluated at the mid-radius of each bin using the best-fit parameters of Table~\ref{tab:bestfit}. The opacity of each track is proportional to the local star formation rate. The infall normalisation is recalibrated for each $v$ to match $M_\star(t_G)=(5.17\pm1.11)\times10^{10}\,M_\odot$
\citep{LicquiaNewman2015}.}
    \label{fig:ofe-feh-flows}
\end{figure*}

\begin{figure*}
    \centering
    \includegraphics[width=0.8\textwidth]{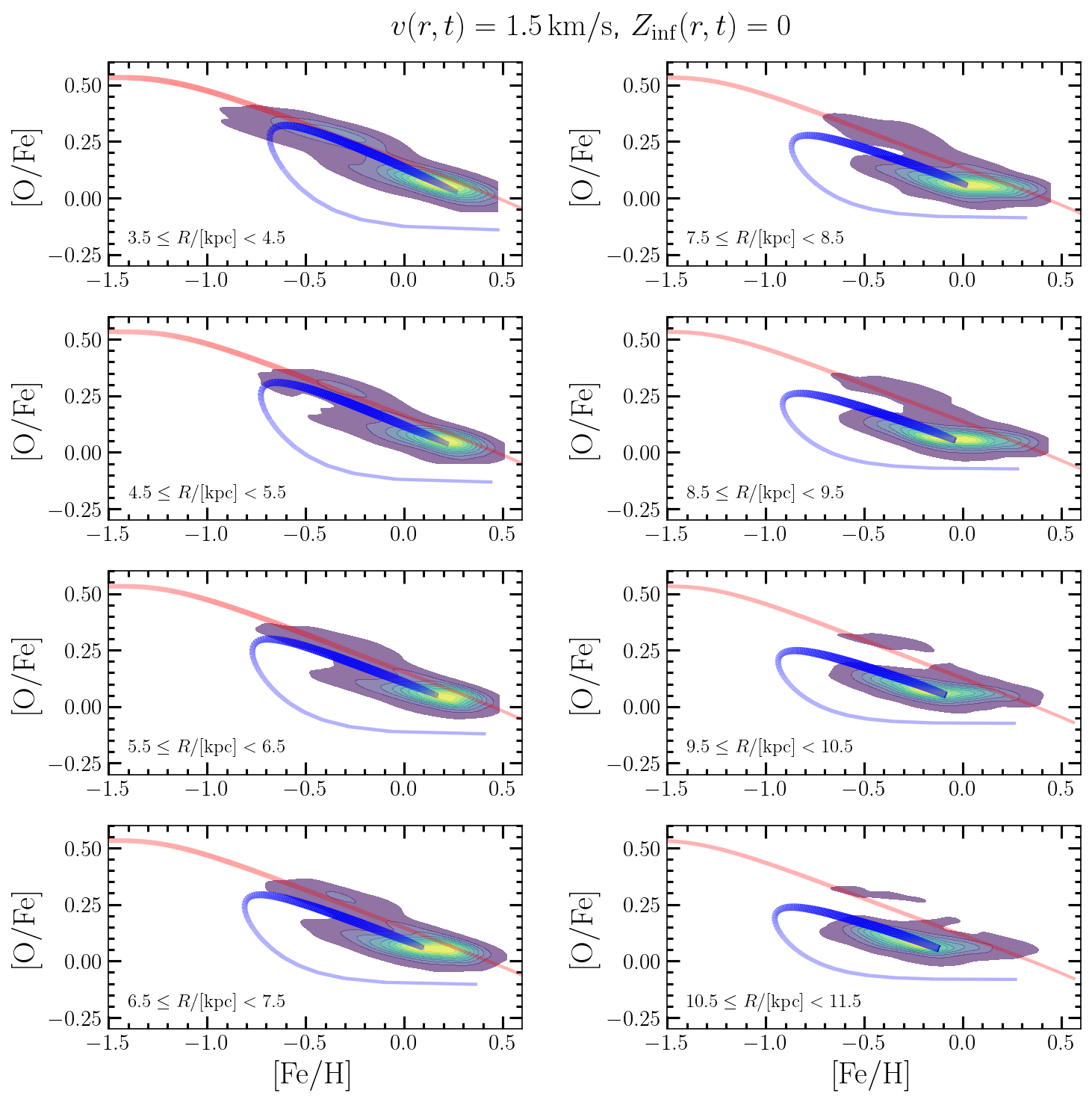}
    \caption{Same as Fig.~\ref{fig:ofe-feh-flows} but showing only the model with constant inflow velocity $v=1.5\,\mathrm{km\,s^{-1}}$. The red and blue curves correspond respectively to the thick-disc ($\tau < 3.8\,\mathrm{Gyr}$) and thin-disc ($\tau \geq 3.8\,\mathrm{Gyr}$) phases, with line width and opacity proportional to the star formation rate.}
    \label{fig:ofe-feh}
\end{figure*}

\section{Fitting procedure}
\label{sec:fitting}
 
\subsection{Free parameters and working assumptions}
\label{sec:params}
 
Throughout the fitting procedure we set $v=0$, so that each annulus evolves independently as a one-zone model (Section~\ref{sec:char}). We also assume that the accreted gas is of primordial composition ($Z_{\rm inf}=0$). The model involves the following parameter vector:

\begin{equation}
\Theta_R = \{\,\tau_1,\ \tau_2,\ t_{\rm max},\ \Sigma_2/\Sigma_1,\
\hat{\epsilon}_{\rm thick},\ \hat{\epsilon}_{\rm thin},\ \hat{\eta}\,\}.
\label{eq:params}
\end{equation}

Three of these are fixed before the fit: \textit{(i)} the transition time, which is set to $t_{\rm max}=3.8$~Gyr; \textit{(ii)} the thin-disc infall timescale, $\tau_2(r)$, which is estimated from the \texttt{astroNN} age distribution of thin-disc stars; \textit{(iii)} the mass-loading factor, which is set to $\hat{\eta}=0$, following the standard assumptions of classical two-infall models of the Milky Way disc \citep[e.g.][]{Spitoni2019,Spitoni2021}. The remaining four parameters $\{\tau_1,\,\Sigma_2/\Sigma_1,\,\hat{\epsilon}_{\rm thick},\,\hat{\epsilon}_{\rm thin}\}$ are treated as free, with the following uniform priors:
\begin{enumerate}
    \item $\tau_1\in(0.1,1.0)$~Gyr;
    \item $\Sigma_2/\Sigma_1\in(3.0,60.0)$;
    \item $\hat{\epsilon}_{\rm thick}\in(0.5,2.5)$~Gyr$^{-1}$;
    \item $\hat{\epsilon}_{\rm thin}\in(0.5,2.5)$~Gyr$^{-1}$.
\end{enumerate}
These four parameters are determined from the fit at $v=0\,\text{km/s}$. Once the best-fit values are established, we explore the effect of a constant radial inflow $v>0\,\text{km/s}$, by re-running the reference model with the same parameters, recalibrating the infall for each $v$ to match the observed MW stellar mass at the present time, from the predicted surface-density profiles
(see Section~\ref{sec:results_stars}).
 
\subsection{Likelihood and best-fit model parameters}
\label{sec:likelihood}

\subsubsection{Likelihood}
 
We fit the model independently in each Galactocentric annulus, covering the range $R=3.5$--$11.5$~kpc in steps of $\Delta R=1.0$~kpc. 

At fixed radius, the observables of the $n$-th star are $x_n=\{[\mathrm{Fe/H}]_n,[\mathrm{O/Fe}]_n\}$ with measurement uncertainties $\sigma_n$. The model prediction is represented by the abundance $\{\mu_i\}$ as given by equation~(\ref{eq:Z_sol}). We note that a defining feature of the two-infall track is that it is \emph{multi-valued} in the [O/Fe]--[Fe/H] plane, because the dilution following the onset of the second infall drives [Fe/H] to lower values, producing a loop.
 
Let $\boldsymbol{x}_n = ([\mathrm{Fe/H}]_n,\,[\mathrm{O/Fe}]_n)$ be the observed abundances of the $n$-th star, with measurement uncertainties $\boldsymbol{\sigma}_n = (\sigma_{n,\mathrm{Fe}},\,\sigma_{n,\mathrm{O}})$, and let $\boldsymbol{m}_i = ([\mathrm{Fe/H}]_i^{\rm mod},\,[\mathrm{O/Fe}]_i^{\rm mod})$ be the model abundances at the $i$-th point along the chemical evolution track. We take the probability of observing the $n$-th star given the model track to be the SFR-weighted sum of Gaussian kernels, as follows:

\begin{equation}
\begin{split}
p_n = \sum_i w_i\, \frac{1}{2\pi\,\sigma_{n,\rm Fe}\,\sigma_{n,\rm O}}\, \exp\Bigg[-\frac{1}{2}\bigg(
&\frac{([\mathrm{Fe/H}]_n - [\mathrm{Fe/H}]_i^{\rm mod})^2}{\sigma_{n,\rm Fe}^2} \\ 
+\ &\frac{([\mathrm{O/Fe}]_n - [\mathrm{O/Fe}]_i^{\rm mod})^2}{\sigma_{n,\rm O}^2} \bigg)\Bigg],
\end{split}
\label{eq:pn}
\end{equation}

where $w_i \propto \psi(t_i)\,\Delta t_i$ is the stellar mass formed at track point $i$, normalised so that $\sum_i w_i = 1$. The objective function minimised in the fit is the total negative log-likelihood, summed over all $N$ stars in the annulus,

\begin{equation}
-2\ln\mathcal{L} = -2\sum_{n=1}^{N} \ln p_n .
\label{eq:loglike}
\end{equation}

The SFR weighting $w_i$ naturally downweights the rapidly-evolving, sparsely-populated phases of the track and concentrates the statistical weight on the well-populated phases. We do not attempt to correct for APOGEE selection effects, so the likelihood serves as an approximate measure of goodness of fit rather than a formally calibrated
probability.

 \begin{figure*}
\centering
\includegraphics[width=0.8\textwidth]{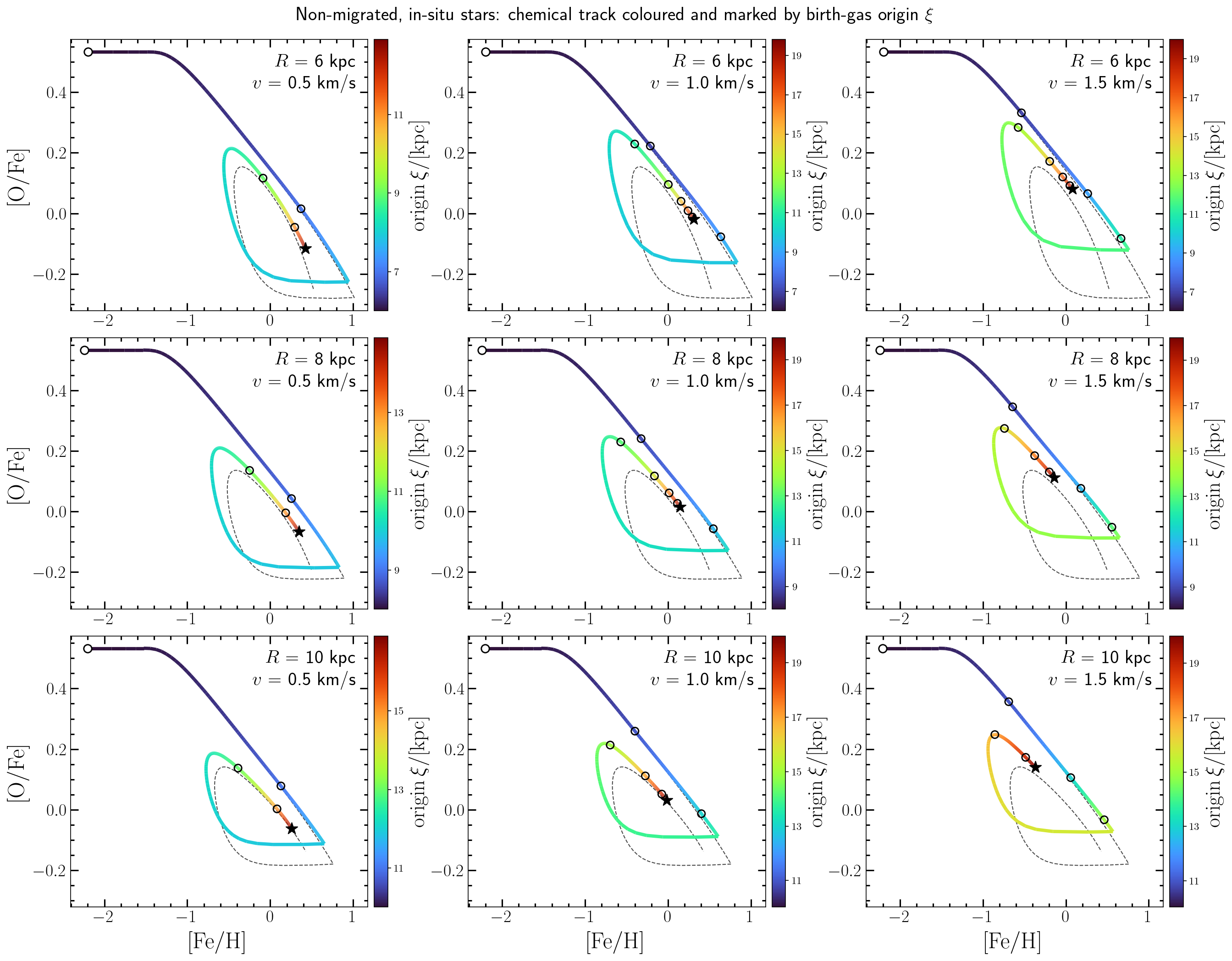}
\caption{Chemical evolution tracks in the [O/Fe]--[Fe/H] plane of the gas residing at fixed Galactocentric radii $R = 6, 8, 10$~kpc (rows) for constant inflow velocities $v = 0.5, 1.0, 1.5~\mathrm{km\,s^{-1}}$ (columns). Each track is the Eulerian abundance history of the gas located at $R$, coloured by the radius $\xi$ from which that gas originated, i.e.\ the Lagrangian label of the characteristic passing through $R$ at each time ($\xi = R + v\,\tau$ for a constant velocity); tracks are truncated at $\xi = 20$~kpc, the outer edge of the modelled disc. Open circles mark steps of 2~kpc in origin $\xi$. The first circle marks the initial valid point at the earliest time ($\xi \simeq R$), and the star symbol marks the present day. The dashed line is the in-situ track ($v = 0$) at the same radius, shown for reference.}
\label{fig:mechanism_stars}
\end{figure*}

\subsubsection{Best-fit model parameters}
 
A global optimum is first located with the differential evolution algorithm \citep{StornPrice1997}, as implemented in \textsc{scipy} \citep{Virtanen2020}, which performs a global search over the prior bounds without requiring an initial guess. The posterior distribution of the free parameters is then sampled with the affine-invariant ensemble sampler of \citet{GoodmanWeare2010}, in its \textsc{emcee} implementation \citep{ForemanMackey2013}, using $18$ walkers initialised around the differential-evolution solution. We use the posterior medians as the best-fit parameters, summarised at different radii in Table~\ref{tab:bestfit}. Since the dominant source of error is systematic rather than statistical, we report only the medians of the marginalised posteriors, without formal uncertainties. 

Beyond the fitted range ($R > 11$~kpc), the radial parameter profiles are held fixed at their values at $R = 11$~kpc. We recall that the thin-disc infall timescale $\tau_2(R)$ is not a fit parameter but is estimated from the \texttt{astroNN} stellar ages (Section~\ref{sec:model}). We verified that varying the extrapolation of $\tau_2(R)$ beyond $11$~kpc by up to $\sim40\%$ leaves the predicted abundance tracks at all fitted radii unchanged.

\begin{figure*}
    \sidecaption
    \includegraphics[width=12cm]{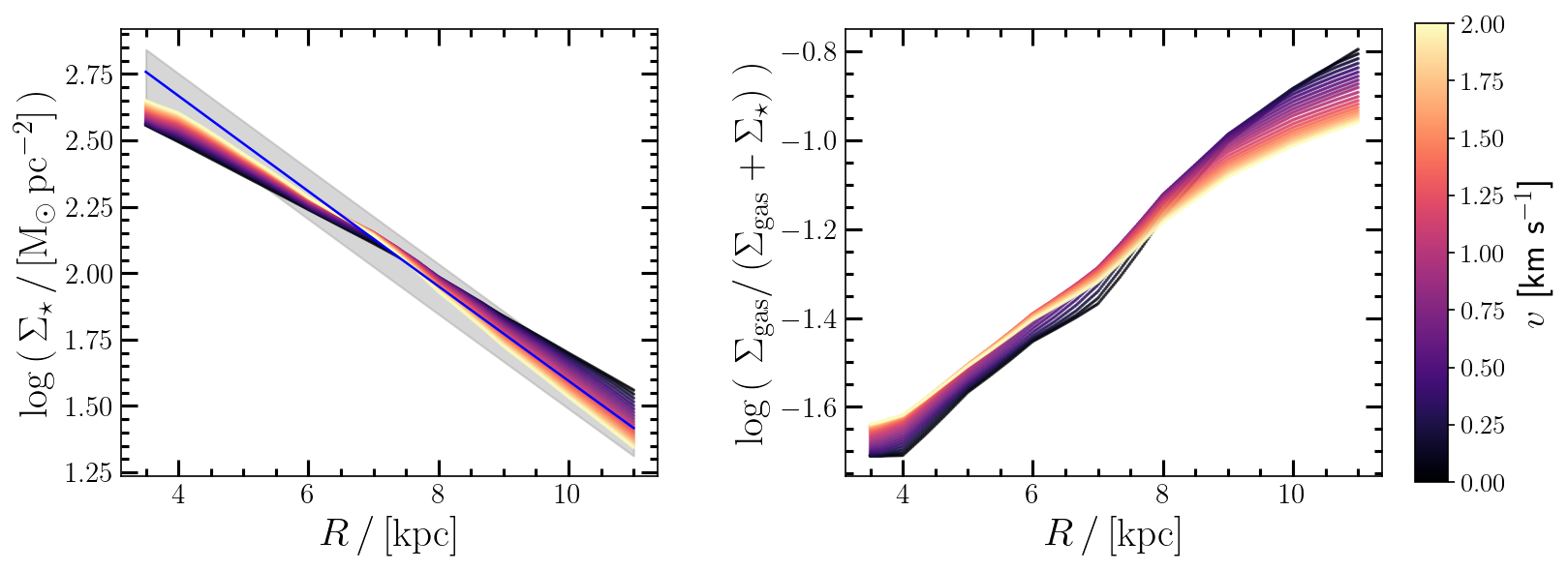}
    \caption{Left: present-day stellar surface-density profile $\Sigma_{\star}(R)$ predicted by the model for inflow velocities $0 \le v \le 2\,\mathrm{km\,s^{-1}}$ (points colour-coded by $v$), compared with the observed Milky Way disc (blue line and grey band), modelled as a double-exponential \citep{BlandHawthorn2016} normalised to the stellar mass $M_{\star}=(5.17\pm1.11)\times10^{10}\,\mathrm{M_{\odot}}$ of \citet{LicquiaNewman2015} over $R \ge 3.5$~kpc; the grey band spans the $\pm1\sigma$ mass uncertainty. Right: corresponding gas fraction $\Sigma_{\mathrm{gas}}/(\Sigma_{\mathrm{gas}}+\Sigma_{\star})$ as a function of radius, for the same velocities.} \label{fig:sigma_star_gas}
\end{figure*}

\begin{figure*}
\centering
\includegraphics[width=\textwidth]{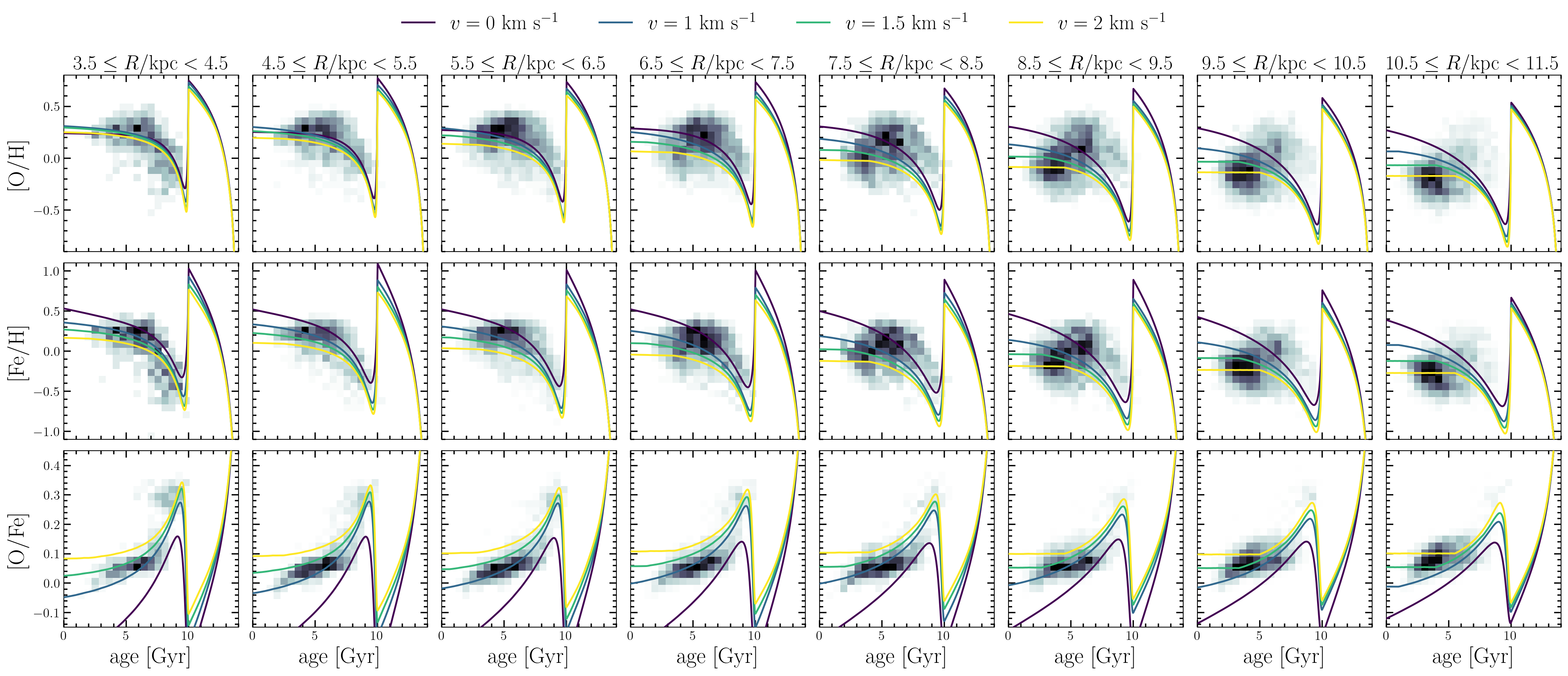}
\caption{Age--abundance relations across the disc. Each column corresponds to a Galactocentric ring (labelled on top), and the three rows show $\mathrm{[O/H]}$, $\mathrm{[Fe/H]}$, and $\mathrm{[O/Fe]}$ as a function of stellar age. The grey two-dimensional histograms show the observed density of the stars in each ring restricted to $|z| <0.5$~kpc, with ages from \texttt{astroNN} \citep{Leung2019,Mackereth2019} and chemical abundances from APOGEE~DR17 \citep{Abdurrouf2022}. The coloured curves are the model tracks assuming the best-fitting parameters for each ring, evaluated at the inflow velocities indicated in the legend.}
\label{fig:age_abundance}
\end{figure*}

\section{Results}
\label{sec:results}
 
After calibrating the model parameters by assuming no radial gas flows ($v=0\,\text{km/s}$) to fit the observed [O/Fe]--[Fe/H] abundance distribution of SDSS-APOGEE DR17 at different Galactocentric distances, in this Section we explore the effect of different constant radial inflows, by re-running the calibrated model for a range of velocities $0 \le v \le 2~\mathrm{km\,s^{-1}}$. The infall rate for each $v$ is rescaled every time to reproduce the present-day stellar mass of the Milky Way. In what follows we present two complementary analyses: the effect of radial gas inflows with different constant velocities on the [O/Fe]--[Fe/H] chemical abundance patterns of stars at a given radius (Section~\ref{sec:results_stars}), and the effect on the radial profiles of gas abundances across the Galactic disc (Section~\ref{sec:results_gas}).
 
\subsection{The effect of radial gas flows on [O/Fe]--[Fe/H]}
\label{sec:results_stars}
 
Figure~\ref{fig:ofe-feh-flows} shows the predictions of our reference model for the [O/Fe]--[Fe/H] abundance patterns at different Galactocentric distances, assuming gas inflow velocities in the range $0 \le v \le 2~\mathrm{km\,s^{-1}}$. The predicted abundance  patterns are computed at the mid-radius of each bin. Increasing the gas inflow velocity does not affect the position of the high-[$\alpha$/Fe] thick-disc stellar populations in the diagram, whereas the low-[$\alpha$/Fe] thin-disc stars are predicted to move towards lower [Fe/H] and higher [O/Fe], improving the agreement with the observed pattern at all radii.
 
With inflow, the gas that today resides at a given radius $R$ originated from a larger radius $\xi = R + v\,t_G$. In our reference model, the outer disc forms fewer stars, for three concurrent reasons: \textit{(i)} the star-formation efficiency decreases with radius over the range $4 \le R \le 11~\mathrm{kpc}$; \textit{(ii)} the thin-disc infall timescale $\tau_2(R)$ increases outward (Table~\ref{tab:tau_thin}), so less gas is available to form stars at any given epoch; \textit{(iii)} the infall normalisation declines with radius, scaling as $\exp[-(r-R_\odot)/r_d]$, so that less gas is accreted cumulatively. As a consequence, the gas now at $R$ was enriched in the past --- while moving inwards --- by the nucleosynthetic products of fewer stars, and in particular by fewer Type~Ia SNe, so that it ends up with lower [Fe/H] and higher [$\alpha$/Fe]. Larger inflow velocities amplify this effect because the gas reaching a given radius originates from increasingly larger radii, where the three effects are stronger. Fig.~\ref{fig:ofe-feh} focuses on the comparison between the observed APOGEE [O/Fe]-[Fe/H] chemical abundance pattern and the model assuming a radial gas inflow with velocity $v \simeq 1.5~\mathrm{km\,s^{-1}}$. In the figure, we separate thin-disc stellar populations in blue from thick-disc stellar populations in red.

When radial gas flows are included, the predicted [O/Fe]--[Fe/H] abundance pattern is still a sequence in stellar age, but the [O/Fe] and [Fe/H] values along the chemical evolution track now reflect different star-formation and chemical-enrichment histories: younger stars form from gas that has flowed in from larger radii, so that --- in the absence of stellar migration --- the youngest in-situ stars at a given radius trace gas assembled in the outer disc, whereas the older thick-disc stars trace gas closer to their present-day radius, which has moved less radially.
This is illustrated in Fig.~\ref{fig:mechanism_stars}, where the [O/Fe]--[Fe/H] abundance patterns at three indicative radii $R=6$, $8$ and $10\,\text{kpc}$ are coloured by the origin $\xi$ of the gas out of which each portion of the track formed. The oldest stars (early $\tau$, with $\xi \simeq R$) formed from locally accreted gas, whereas the youngest stars (towards the present day, with larger $\xi$) formed from gas that had flowed in from the outer disc. A faster inflow shifts the birth-gas origin of the young in-situ populations to progressively larger $\xi$, lowering their [Fe/H] and raising their [O/Fe].

The same mechanism is reflected in the present-day stellar surface-density profile (Fig.~\ref{fig:sigma_star_gas}). Radial gas flows redistribute gas inward, to regions where the star-formation efficiency is higher; because the star-formation rate is assumed proportional to the gas surface mass density, this causes a steepening of $\Sigma_\star(R)$. All reference models yield a gas fraction that increases outward, as expected for an inside-out forming disc.

Figure~\ref{fig:age_abundance} shows the age--abundance relations predicted by our reference chemical evolution models with different gas flow velocities across the disc, compared with the observed values from APOGEE and \texttt{astroNN} in each ring. All tracks display a sharp drop around an age of $\sim9$--$10$~Gyr, characteristic of the two-infall scenario. In $\mathrm{[O/H]}$ (top row) and $\mathrm{[Fe/H]}$ (middle row), the tracks with higher inflow velocity in the thin-disc phase lie systematically below those at $v=0$, whereas in $\mathrm{[O/Fe]}$ (bottom row) the tracks with higher inflow velocity lie systematically above those at $v=0$, mirroring the behaviour seen in the $\mathrm{[O/Fe]}$--$\mathrm{[Fe/H]}$ plane and improving the agreement with the observed trends as a function of stellar age.
 
\subsection{Radial abundance gradients}
\label{sec:results_gas}

\begin{table*}
\centering
\caption{Radial abundance gradient as predicted by our models at Galactocentric distances in the range $6 \le R_{\rm GC} \le 16$~kpc, for different inflow velocities. }
\label{tab:gradient_inner}
\begin{tabular}{lccccc}
\hline\hline
 & \multicolumn{4}{c}{Model $v$ [$\mathrm{km\,s^{-1}}$]} & OCCAM \\ Age bin [Gyr] & $0.0$ & $1.0$ & $1.5$ & $2.0$ & DR17 \\
\hline
\multicolumn{6}{l}{$\mathrm{d[Fe/H]}/\mathrm{d}R$}\\
$\le 0.4$     & $-0.017$ & $-0.044$ & $-0.052$ & $-0.055$ & $-0.052\pm0.003$ \\
$0.4$--$0.8$  & $-0.017$ & $-0.042$ & $-0.052$ & $-0.054$ & $-0.059\pm0.003$ \\
$0.8$--$2.0$  & $-0.018$ & $-0.040$ & $-0.052$ & $-0.054$ & $-0.059\pm0.002$ \\
$2.0$--$6.0$  & $-0.022$ & $-0.033$ & $-0.048$ & $-0.054$ & $-0.052\pm0.002$ \\
All (present) & $-0.016$ & $-0.044$ & $-0.052$ & $-0.055$ & $-0.055\pm0.001$ \\
\hline
\multicolumn{6}{l}{$\mathrm{d[O/H]}/\mathrm{d}R$}\\
$\le 0.4$     & $-0.004$ & $-0.037$ & $-0.044$ & $-0.047$ & \dots \\
$0.4$--$0.8$  & $-0.005$ & $-0.036$ & $-0.044$ & $-0.047$ & \dots \\
$0.8$--$2.0$  & $-0.007$ & $-0.035$ & $-0.044$ & $-0.047$ & \dots \\
$2.0$--$6.0$  & $-0.012$ & $-0.030$ & $-0.042$ & $-0.046$ & \dots \\
All (present) & $-0.003$ & $-0.037$ & $-0.044$ & $-0.047$ & \dots \\
\hline
\end{tabular}
\tablefoot{For comparison, we report the measured gradients of [Fe/H] in a sample of open clusters of the OCCAM survey \citep{Myers2022}. Model gradients are measured on the Eulerian gas-phase abundance profile evaluated at the representative formation time $t = t_G - \mathrm{age_{med}}$ of each age bin, where $\mathrm{age_{med}}$ is the median age of the OCCAM clusters in that bin (and at the present day for the full sample). The fitting range matches the radial extent of the OCCAM clusters. The reported OCCAM $\mathrm{[Fe/H]}$ gradients correspond to the single linear fits as reported in the original work; the observed $\mathrm{[O/H]}$ gradient is not present because it is not tabulated by OCCAM in age bins. All values are in $\mathrm{dex\,kpc^{-1}}$.}
\end{table*}
 
\begin{figure*}
    \centering
    \includegraphics[width=0.8\textwidth]{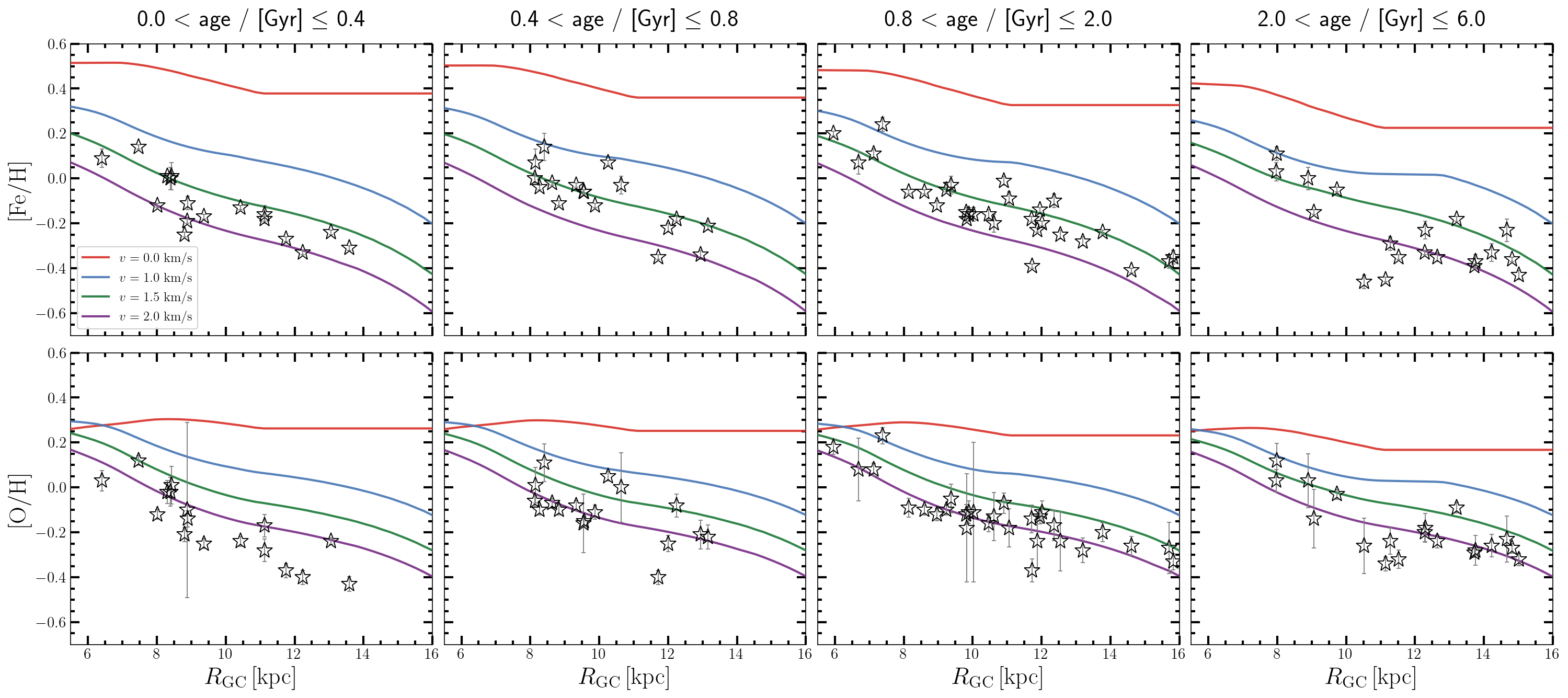}
    \caption{Radial gradients of $[\mathrm{Fe/H}]$ (top row) and $[\mathrm{O/H}]$ (bottom row) for the OCCAM sample of open clusters \citep{Myers2022}, divided into the same four age bins used in their analysis (black stars with error bars; for $[\mathrm{O/H}]$ the uncertainty is propagated from $[\mathrm{Fe/H}]$ and $[\mathrm{O/Fe}]$). In each panel the coloured lines show the predicted Eulerian gas-phase abundance profiles of models for constant inflow velocities $v=0$, $1.0$, $1.5$, and $2.0\,\mathrm{km\,s^{-1}}$, evaluated at the representative formation time $t=t_G-\mathrm{age_{med}}$ corresponding to the median cluster age of the bin (indicated above each column).}
    \label{fig:gradient-agebins}
\end{figure*}

In Fig.~\ref{fig:gradient-agebins}, we show how [Fe/H] (top row) and [O/H] (bottom row) are predicted to change as a function of Galactocentric distance for different radial gas inflow velocities (different curves) and in different age bins (different columns). The predictions of our reference models are compared with observations in a sample of open clusters of the OCCAM survey \citep{Myers2022}. In Table~\ref{tab:gradient_inner} we report the predicted slopes of the [Fe/H] and [O/H] radial profiles, measured over the same radial range spanned by the OCCAM clusters ($6 \le R_{\rm GC} \le 16$~kpc), reporting also the measured slopes of [Fe/H] in the observed sample of \citet{Myers2022}.
Without radial flows, ($v=0$) the predicted gradients of [Fe/H] and [O/H] are almost flat in all age bins, whereas models with radial gas inflows of velocity $v \simeq 1.5~\mathrm{km\,s^{-1}}$ provide a better agreement with the observed slopes across all age bins. This is consistent with the finding that inside-out growth alone produces too shallow a gradient, and radial gas flows are required to recover the observed slopes of the radial profiles of the abundances \citep{Spitoni2011, Bilitewski2012}. The gradient steepens monotonically with $v$ at all stellar ages; the absolute [Fe/H] and [O/H] abundances continue to decrease with $v$ at all radii, providing a complementary constraint. 
These trends are in agreement with those of previous chemical evolution models, which found that the effect of radial gas inflows is to steepen the abundance gradients while lowering the absolute abundances \citep{Spitoni2013, Palla2020, Johnson2025}.
 
 \begin{figure*}
\centering
\includegraphics[width=0.8\textwidth]{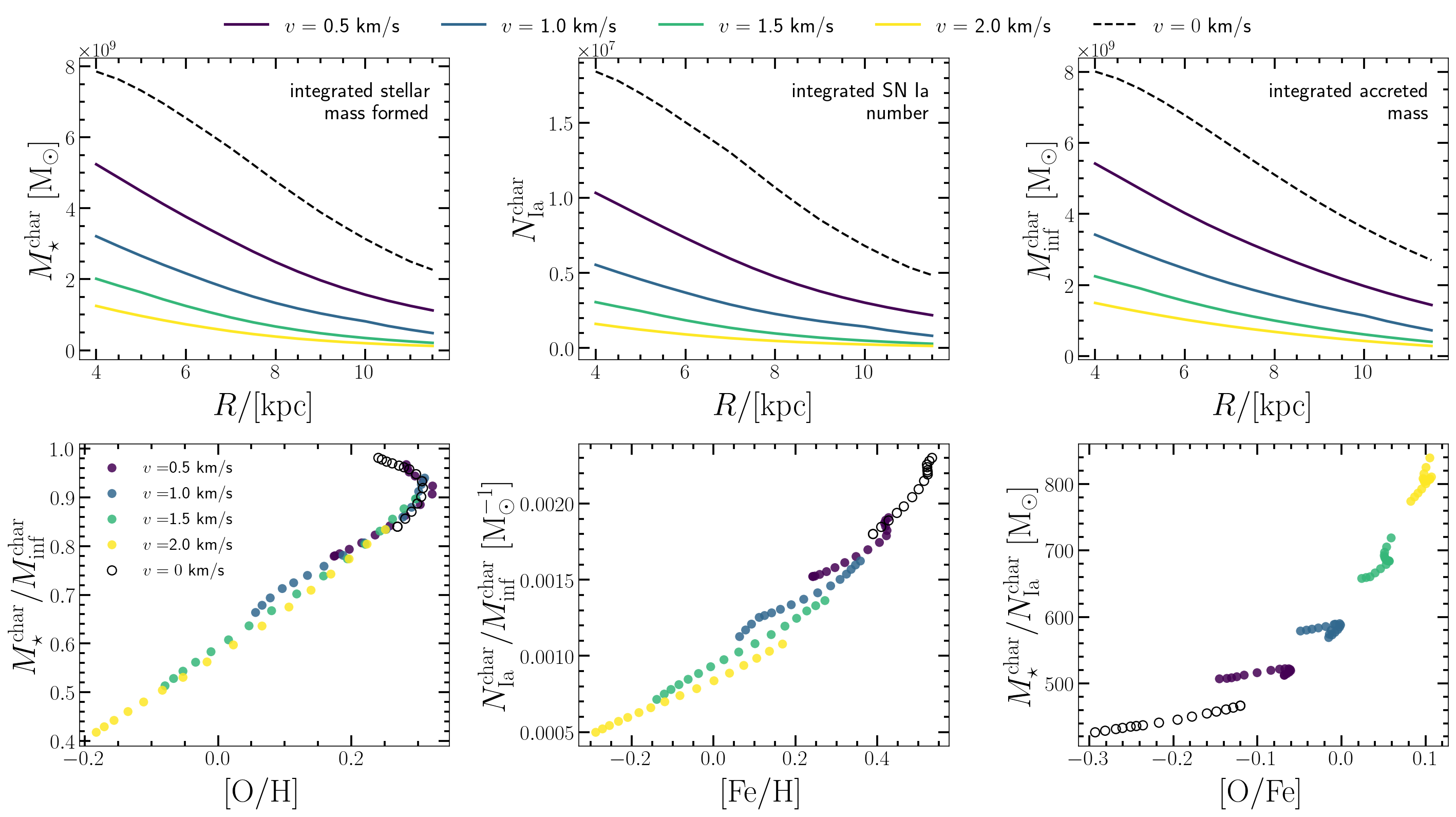}
\caption{\emph{Top row:} the total stellar mass formed $M_\star^{\rm char}$, the total number of Type~Ia SNe $N_{\rm Ia}^{\rm char}$, and the total accreted gas mass $M_{\rm inf}^{\rm char}$, integrated along the characteristic of the gas that reaches each radius $R$ today (equations~\ref{eq:integrated}), as a function of $R$, for inflow velocities $v=0.5,\,1.0,\,1.5,\,2.0~\mathrm{km\,s^{-1}}$ (coloured solid lines) and for the in-situ case with $v=0$ (dashed line). \emph{Bottom row:} the present-day gas abundances against the ratios that drive them; from left to right, [O/H] versus $M_\star^{\rm char}/M_{\rm inf}^{\rm char}$, [Fe/H] versus $N_{\rm Ia}^{\rm char}/M_{\rm inf}^{\rm char}$, and [O/Fe] versus $M_\star^{\rm char}/N_{\rm Ia}^{\rm char}$. Different colours correspond to different inflow velocities $v$, with the in-situ case ($v=0$) being shown using open circles.}
\label{fig:mechanism_gas}
\end{figure*}

\begin{figure*}
    \centering
    \includegraphics[width=0.8\textwidth]{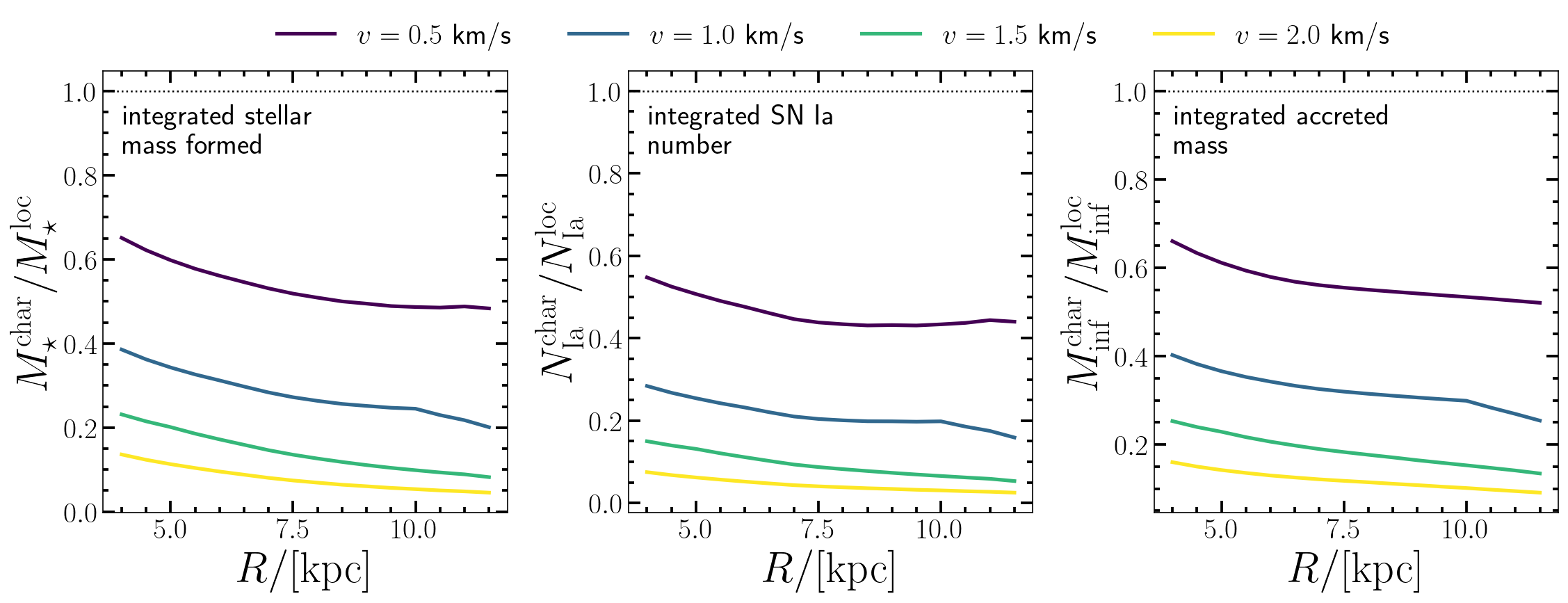}
    \caption{Ratio between the quantities integrated along the flow characteristic, $M_\star^{\rm char}$, $N_{\rm Ia}^{\rm char}$, and $M_{\rm inf}^{\rm char}$, and the corresponding quantities that would be measured locally at the same radius, $M_\star^{\rm loc}$, $N_{\rm Ia}^{\rm loc}$, and $M_{\rm inf}^{\rm loc}$, as a function of Galactocentric radius $R$, for inflow velocities $v=0.5,\,1.0,\,1.5,\,2.0~\mathrm{km\,s^{-1}}$. The dotted line marks unity, where the quantities on the characteristics would coincide with those locally.}
    \label{fig:lagrangian_vs_eulerian}
\end{figure*}
 
\begin{figure*}
    \centering
    \includegraphics[width=0.8\textwidth]{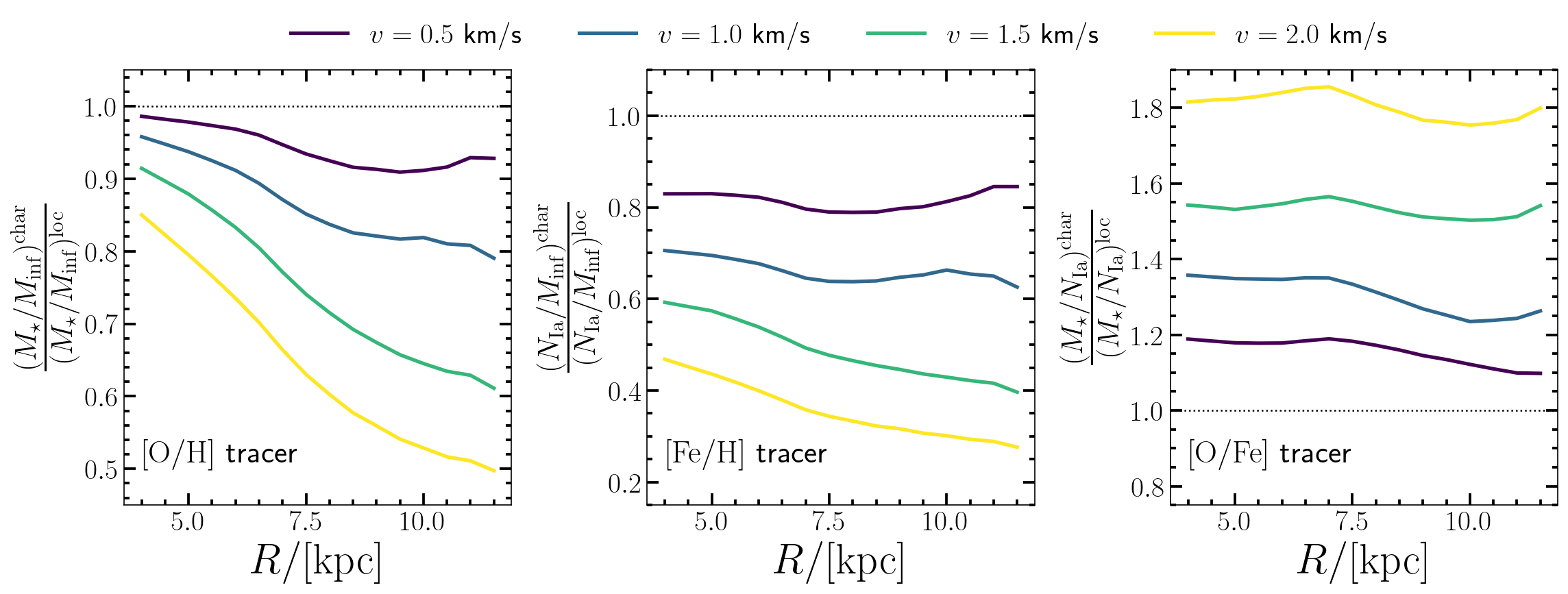}
    \caption{Effect of radial gas flows on the ratios that drive the abundances, shown as the ratio between each driver evaluated along the flow characteristic and the same driver that would be measured locally at the same radius, as a function of Galactocentric radius $R$, for inflow velocities $v=0.5,\,1.0,\,1.5,\,2.0~\mathrm{km\,s^{-1}}$. From left to right: $M_\star/M_{\rm inf}$, $N_{\rm Ia}/M_{\rm inf}$, and $M_\star/N_{\rm Ia}$. The dotted line marks unity, where the characteristic and local values would coincide.}
    \label{fig:ratios_char_vs_loc}
\end{figure*}
 
To quantify the enrichment history experienced by the gas reaching a given radius today, we follow the characteristic crossing the radius $R$ at the present time and integrate the relevant rates along its trajectory. The results of our analysis are shown in Fig.~\ref{fig:mechanism_gas}. The top panel on the left shows the total stellar mass formed, $M_\star^{\rm char}$, from the gas parcels while they flow inwards along the characteristics until they reach their present-day radius; this quantity directly reflects the cumulative star-formation activity experienced by the gas along the different characteristics ending up at different $R$ today. The top panel in the middle shows the integrated number of SNe Ia that have enriched the same gas parcels, $N_{\rm Ia}^{\rm char}$, and the top panel on the right shows the total accreted gas mass along the characteristic, $M_{\rm inf}^{\rm char}$. The bottom row shows how the predicted [O/H] (left panel), [Fe/H] (middle panel) and [O/Fe] (right panel) in the Galaxy disc at the present time, correlate with the ratios $M_\star^{\rm char}/M_{\rm inf}^{\rm char}$, $N_{\rm Ia}^{\rm char}/M_{\rm inf}^{\rm char}$ and $M_\star^{\rm char}/N_{\rm Ia}^{\rm char}$, respectively; the predictions for different gas inflow velocities are shown using different colours. The predictions of the models without radial gas inflows are shown as open circles in the bottom row and as a dashed line in the top row.
 
Since the star-formation rate $\psi$, the Type~Ia rate $R_{\rm Ia}$, and the infall rate $\hat{\mathcal{I}}$ are surface densities, to make Fig.~\ref{fig:mechanism_gas} each of them is weighted by the physical area of the gas parcel as it flows inward. For a constant inflow velocity the characteristics are parallel, so that a parcel of fixed Lagrangian width $\Delta\xi$ retains the same width at all times, and the area of the annulus it occupies at time $\tau$ is given by $2\pi\,r(\tau)\,\Delta\xi$, where $r(\tau)$ is the Galactocentric radius of the parcel at time $\tau \le t_G$. The three quantities $M_\star^{\rm char}$, $N_{\rm Ia}^{\rm char}$, and $M_{\rm inf}^{\rm char}$ are then given by
\begin{subequations}
\label{eq:integrated}
\begin{align}
M_\star^{\rm char}    &= (1-\hat{R})\int_0^{t_G} \psi\,[2\pi\,r(\tau)\,\Delta\xi]\,\mathrm{d}\tau, \label{eq:Mstar}\\
N_{\rm Ia}^{\rm char} &= \int_0^{t_G} R_{\rm Ia}\,[2\pi\,r(\tau)\,\Delta\xi]\,\mathrm{d}\tau, \label{eq:NIa}\\
M_{\rm inf}^{\rm char} &= \int_0^{t_G} \hat{\mathcal{I}}\,[2\pi\,r(\tau)\,\Delta\xi]\,\mathrm{d}\tau, \label{eq:Minf}
\end{align}
\end{subequations}
where $R_{\text{Ia}}$ is the SN~Ia rate and $\hat{R}$ is the return fraction (Section~\ref{sec:starformation}). We adopt a common Lagrangian width $\Delta\xi=1$~kpc for all radii and velocities, so that the curves in the top row of Fig.~\ref{fig:mechanism_gas} can be directly compared.
 
The top row of Fig.~\ref{fig:mechanism_gas} shows that $M_\star^{\rm char}$, $N_{\rm Ia}^{\rm char}$, and $M_{\rm inf}^{\rm char}$ all decline towards larger radii and are further reduced when assuming radial gas inflows with increasing velocity, reflecting the fact that the gas reaching a given radius is increasingly drawn from the outer disc --- with lower star-formation efficiency, longer infall timescales, and lower infall normalisation --- as $v$ increases. In the absence of radial flows ($v=0$, dashed lines) these quantities measure the local star formation and chemical enrichment history at each radius and lie systematically above the quantities with $v > 0\,\text{km}\,\text{s}^{-1}$.

The bottom panels of Fig.~\ref{fig:mechanism_gas} isolate how [O/H], [Fe/H], and [O/Fe] correlate with the ratios of the integrated quantities shown in the top panels. The present-day [O/H] abundances, for radial gas inflow velocities $v \gtrsim 1.0~\mathrm{km\,s^{-1}}$, show a positive correlation with the surviving stellar mass per unit accreted gas mass, $M_\star^{\rm char}/M_{\rm inf}^{\rm char}$; for lower inflow velocities, a positive correlation is seen only in the inner disc, turning into an anti-correlation in the outer disc, which indicates that the integrated accreted gas mass increases more rapidly than the integrated stellar mass formed along those characteristics. This turnover moves inward from $R\simeq8.5$~kpc for $v=0~\mathrm{km\,s^{-1}}$ to $R\simeq6.5$~kpc for $v=0.5~\mathrm{km\,s^{-1}}$. The present-day [Fe/H] abundances show a positive correlation with the number of Type~Ia SNe per unit accreted gas mass, $N_{\rm Ia}^{\rm char}/M_{\rm inf}^{\rm char}$, for all radial inflow velocities considered. Finally, [O/Fe] correlates with the stellar mass formed per Type~Ia SN, $M_\star^{\rm char}/N_{\rm Ia}^{\rm char}$, clearly separating the different inflow velocities. As the inflow velocity increases, the integrated number of Type~Ia SNe decreases more steeply than the integrated stellar mass, so that $M_\star^{\rm char}/N_{\rm Ia}^{\rm char}$ rises and [O/Fe] increases, on average, with $v$.

In Fig.~\ref{fig:lagrangian_vs_eulerian} we quantify the difference between the star formation, gas accretion and Type Ia SN chemical enrichment histories experienced by the gas along its inward trajectory and the local history at each radius. This is obtained by comparing the integrated quantities along the characteristic, $M_\star^{\rm char}$, $N_{\rm Ia}^{\rm char}$, and $M_{\rm inf}^{\rm char}$, with the corresponding quantities $M_\star^{\rm loc}$, $N_{\rm Ia}^{\rm loc}$, and $M_{\rm inf}^{\rm loc}$ for the local history at the same radius. We find that the gas reaching a given radius has formed fewer stars, produced fewer Type~Ia SNe, and accreted less gas than what would be inferred by treating that radius as an independent one-zone model. At $v\simeq1.5~\mathrm{km\,s^{-1}}$ the integrated star formation experienced by the gas is only $\sim10$--$20$ per cent of the local value across the disc, and the discrepancy grows further at higher velocities. This shows that, once radial flows are present, the present-day abundances at a given radius cannot be attributed to the local star-formation and accretion history at that radius, and that following the gas along its trajectory is essential to recover the correct enrichment history.
 
The effect of the flows on the abundances can be understood by examining how they alter the three ratios, $M_\star/M_{\rm inf}$, $N_{\rm Ia}/M_{\rm inf}$, and $M_\star/N_{\rm Ia}$, which set [O/H], [Fe/H], and [O/Fe] respectively. Fig.~\ref{fig:ratios_char_vs_loc} shows, for each ratio, its value along the characteristic divided by its value for the local history at the same radius, so that a value of unity would mean that radial flows leave the ratio unchanged. The ratios involving $M_\star/M_{\rm inf}$ (tracer of [O/H]) and $N_{\rm Ia}/M_{\rm inf}$ (tracer of [Fe/H]) both lie below unity and decrease with the inflow velocity, reaching $\sim0.5$ and $\sim0.3$ respectively at $v=2~\mathrm{km\,s^{-1}}$; this means that the gas reaching a given radius has formed fewer stars and produced fewer Type~Ia SNe per unit accreted gas than what would be measured locally, lowering [O/H] and, more strongly, [Fe/H]. The ratio involving $M_\star/N_{\rm Ia}$ lies above unity and increases with the inflow velocity, reaching $\sim1.8$ at $v=2~\mathrm{km\,s^{-1}}$; because the Type~Ia SN rate is delayed, the gas arriving from the outer disc has formed more stars per Type~Ia SN than what would be measured locally, raising [O/Fe]. This opposite behaviour of $M_\star/N_{\rm Ia}$ explains why radial flows lower the overall metallicity while amplifying the $\alpha$-enhancement, making [O/Fe] a sensitive tracer of the inflow velocity in this model.

\section{Conclusions and discussion}
\label{sec:conclusions}

We have presented explicit integral solutions for the chemical evolution of a galactic disc threaded by radial gas flows. For an arbitrary radial velocity field $v(r,t)$ and accretion history $\hat{\mathcal{I}}(r,t)$, the model follows, in a Lagrangian frame and along the same flow characteristics, the time evolution of the gas surface mass density and of the abundances of $\alpha$-elements (here we focus on oxygen) and iron, assuming the instantaneous recycling approximation for the core-collapse SN chemical enrichment, delayed Type~Ia SN chemical enrichment calculated along the flow characteristics, and the possibility of an enriched infall. The solutions require only one-dimensional numerical integrals, so that a complete model is fast to evaluate and can be embedded in a likelihood analysis. 

We applied the model to the Milky Way disc, adopting a two-infall accretion history and fitting the [O/Fe]--[Fe/H] abundance distribution of SDSS-APOGEE DR17 at different annuli assuming $v=0$. Keeping the best-fit parameters of the two-infall model fixed, we then used the general $v\neq0$ solution to assess the impact of a constant radial inflow on the disc properties. While our explicit integral solutions are of general validity and are flexible enough that they can be changed to explore a variety of different models and assumptions for the chemical evolution of the Milky Way, the findings of our application are only valid for a two-infall model assuming instantaneous recycling for the chemical enrichment of core-collapse SNe and delayed chemical enrichment of Type Ia SNe calculated along the flow characteristics, being not directly transferable to generic Galactic chemical evolution models, especially multi-zone or even chemodynamical models. We also caution that, in our application, we assume a SFE that can arbitrarily change between the first and the second infall episode \citep{Chiappini1997,Spitoni2020}, introducing two separate parameters that can also individually vary with radius \citep{Spitoni2021}. With these caveats in mind, our main results are the following.
 
\begin{enumerate}
\item A constant inflow velocity $v \simeq 1.5~\mathrm{km\,s^{-1}}$ can reproduce simultaneously the observed [O/Fe]--[Fe/H] abundance patterns across the disc as measured in a sample of red giants by SDSS-APOGEE DR17 (Figs~\ref{fig:ofe-feh-flows} and \ref{fig:ofe-feh}), the present-day stellar surface-density profile of the Milky Way from \citet{BlandHawthorn2016} (Fig.~\ref{fig:sigma_star_gas}), and the radial abundance gradients of [Fe/H] and [O/H] as measured in a sample of open clusters from the OCCAM survey \citep{Myers2022} (Fig.~\ref{fig:gradient-agebins}; Table~\ref{tab:gradient_inner}). The required velocity is small, consistent with the low inflow velocities inferred for the Milky Way disc by previous studies \citep{Bilitewski2012, Pezzulli2016, Johnson2025}.
 
\item Without radial gas flows the predicted abundance gradients are essentially flat for our best-fit parameters. The slope of the radial abundance gradients steepens monotonically with $v$; the absolute oxygen and iron abundances continue to decrease with $v$, providing a complementary constraint on the radial gas inflow velocity. This is in line, for example, with \citet{Johnson2025}, who reach the same qualitative conclusion from independent multi-zone models.
 
\item When radial gas flows are present, the chemical abundances observed in the gas at a given radius carry the imprint of the star formation and accretion experienced in the past by the same gas along its inward trajectory. For our Milky Way model, the integrated star formation experienced by the gas reaching a given radius is only $\sim10$--$20$ per cent of the local value at $v\simeq1.5~\mathrm{km\,s^{-1}}$, and the discrepancy grows with the assumed inflow velocity (Fig.~\ref{fig:lagrangian_vs_eulerian}). Treating each radius as an independent one-zone model would therefore misrepresent the enrichment history of the gas.
 
\item The abundance gradient arises because the inflow transports gas inward from the outer disc, where less star formation has taken place, for three concurrent reasons: the star-formation efficiency decreases with radius, the infall timescale lengthens, and the infall normalisation declines as $\exp[-(r-R_\odot)/r_d]$ (Fig.~\ref{fig:mechanism_gas}). The present-day [O/H] and [Fe/H] are each set by the relevant nucleosynthetic source --- prompt core-collapse SN production for oxygen, delayed Type~Ia SN production for iron --- diluted by the gas accreted along the trajectory.
 
\item The stellar mass formed per Type~Ia SN is higher by up to $\sim50$ per cent at $v=1.5~\mathrm{km\,s^{-1}}$ relative to what would be measured locally (Fig.~\ref{fig:ratios_char_vs_loc}), because the delayed Type~Ia SN enrichment lags behind the star formation in the gas arriving from larger radii. This mechanism raises [O/Fe], making the $\alpha$-enhancement a sensitive diagnostic of the inflow velocity, complementary to the gradient slope.

 \item Higher inflow velocities in the thin-disc phase lower [O/H] and [Fe/H] and raise [O/Fe] at fixed stellar age, improving the agreement with the observed age--abundance relations across the disc from APOGEE and \texttt{astroNN} (Fig.~\ref{fig:age_abundance}).
 
\item Because the model transports gas before star formation takes place at a given radius, without including stellar radial migrations, it isolates the chemical signature of radial gas flows from the dynamical mixing of already-formed stars (Fig.~\ref{fig:mechanism_stars}). In this picture, the older in-situ stars at any radius formed from locally accreted gas, whereas the younger populations formed from gas that flowed in from larger radii.
\end{enumerate}
 
Coupling the present gas-flow solutions to a treatment of stellar radial migration, and applying our chemical evolution model to time-dependent and/or radially varying velocity fields, are natural extensions of our work, potentially gauged to cosmological simulations. The numerical code is made publicly available.
 
\begin{acknowledgements}
I thank the anonymous referee for all their comments and questions, which greatly improved the quality and clarity of the paper. I am grateful to James Johnson and David Weinberg for several fruitful discussions on the topics of this work. 
\end{acknowledgements}
 
\section*{Data availability}
The numerical code implementing the model presented in this work, together with the scripts used to reproduce all the figures, is publicly available on GitHub at \url{https://github.com/fiorewin/lagrangian-disc-chemical-evolution}. This work is based on publicly available data from SDSS IV APOGEE DR~17 \citep{Abdurrouf2022}, that can be downloaded at \url{https://www.sdss4.org/dr17}. The OCCAM sample of open clusters \citep{Myers2022}, and the \texttt{astroNN} catalogue of stellar ages and Galactocentric positions \citep{Leung2019, Mackereth2019}  are also publicly available on the SDSS IV website as value-added catalogues for APOGEE DR17.

\bibliographystyle{aa}
\bibliography{references}
 
\end{document}